\documentclass[article]{aa}  

\usepackage{graphicx}
\usepackage{amsmath}
\usepackage{verbatim}
\usepackage{fancyvrb}
\usepackage{txfonts}
\usepackage{hyperref}
\usepackage{natbib}
\bibpunct{(}{)}{;}{a}{}{,} 
\usepackage{color}
\usepackage{xspace}
\usepackage{listings}

\definecolor{codegreen}{rgb}{0,0.6,0}
\definecolor{codegray}{rgb}{0.5,0.5,0.5}
\definecolor{codepurple}{rgb}{0.58,0,0.82}
\definecolor{backcolour}{rgb}{0.95,0.95,0.92}

\lstdefinestyle{mystyle}{
    backgroundcolor=\color{backcolour},   
    commentstyle=\color{codegreen},
    keywordstyle=\color{magenta},
    numberstyle=\tiny\color{codegray},
    stringstyle=\color{codepurple},
    basicstyle=\ttfamily\footnotesize,
    breakatwhitespace=false,         
    breaklines=true,                 
    captionpos=b,                    
    keepspaces=true,                 
    numbers=left,                    
    numbersep=5pt,                  
    showspaces=false,                
    showstringspaces=false,
    showtabs=false,                  
    tabsize=2
}

\newcommand{\Gaia}{\textit{Gaia}\xspace}
\newcommand{\ruwe}{\texttt{ruwe}\xspace}

\begin{document}

\title{\Gaia DR3 detectability of unresolved binary systems}

\author{
    Alfred Castro-Ginard          \inst{\ref{inst:leiden}}\relax
\and  Zephyr Penoyre \inst{\ref{inst:leiden}}\relax
\and Andrew R. Casey \inst{\ref{inst:monash},\ref{inst:astro3d}}\relax
\and Anthony G.A. Brown \inst{\ref{inst:leiden}}\relax
\and Vasily Belokurov \inst{\ref{inst:cambridge}}\relax
\and Tristan Cantat-Gaudin \inst{\ref{inst:heidelberg}}\relax
\and Ronald Drimmel \inst{\ref{inst:torino}}\relax
\and Morgan Fouesneau \inst{\ref{inst:heidelberg}}\relax
\and Shourya Khanna \inst{\ref{inst:torino}}\relax
\and Evgeny P. Kurbatov \inst{\ref{inst:cambridge}}\relax
\and Adrian M. Price-Whelan \inst{\ref{inst:flatiron}}\relax
\and Hans-Walter Rix \inst{\ref{inst:heidelberg}}\relax
\and Richard L. Smart \inst{\ref{inst:torino}}
}

\institute{Leiden Observatory, Leiden University, Einsteinweg 55, 2333 CC Leiden, the Netherlands\\ \email{acastro@strw.leidenuniv.nl}\relax \label{inst:leiden}
\and{School of Physics and Astronomy, Monash University, VIC 3800, Australia \label{inst:monash}}
\and{Centre of Excellence for Astrophysics in Three Dimensions (ASTRO-3D), Melbourne, Victoria, Australia \label{inst:astro3d}}
\and{Institute of Astronomy, University of Cambridge, Madingley Road, Cambridge CB3 0HA, United Kingdom \label{inst:cambridge}}
\and{Max-Planck-Institut f\"ur Astronomie, K\"onigstuhl 17, D-69117 Heidelberg, Germany \label{inst:heidelberg}}
\and{INAF - Osservatorio Astrofisico di Torino, Strada Osservatorio 20, Pino Torinese 10025 Torino, Italy \label{inst:torino}}
\and{Center for Computational Astrophysics, Flatiron Institute, 162 Fifth Ave, New York, NY 10010, USA \label{inst:flatiron}}
}

\date{Received date /
Accepted date}

\date{Received date /
Accepted date}
  
\abstract{
    \Gaia can not individually resolve very close binary systems, however, the collected data can still be used to identify them. A powerful indicator of stellar multiplicity is the sources reported Renormalized Unit Weight Error (\ruwe), which effectively captures the astrometric deviations from single-source solutions. 
}{
    We aim to characterise the imprints left on \ruwe caused by binarity. By flagging potential binary systems based on \ruwe, we aim to characterise which of their properties will contribute the most to their detectability.
}{
    We develop a model to estimate \ruwe values for observations of \Gaia sources, based on the biases to the single-source astrometric track arising from the presence of an unseen companion. Then, using the recipes from previous GaiaUnlimited selection functions, we estimate the selection probability of sources with high \ruwe, and discuss what binary properties contribute to increasing the sources' \ruwe.
}{
    We compute the maximum \ruwe value which is compatible with single-source solutions as a function of their location on-sky. We see that binary systems selected as sources with a \ruwe higher than this sky-varying threshold have a strong detectability window in their orbital period distribution, which peaks at periods equal to the \Gaia observation time baseline.
}{
    We demonstrate how our sky-varying \ruwe threshold provides a more complete sample of binary systems when compared to single sky-averaged values by studying the unresolved binary population in the \Gaia Catalogue of Nearby Stars. We provide the code and tools used in this study, as well as the sky-varying \ruwe threshold through the GaiaUnlimited Python package.
}
\maketitle
%


\section{Introduction}
\label{sec:intro}

A large number of the stars in our Galaxy are formed in binary or multiple systems. The fraction of stars hosting at least one companion depends on the star's properties, e.g., for main sequence (MS) stars this fraction increases with stellar mass \citep{2013ARA&A..51..269D}, with $\lesssim 50\%$ of solar-type stars being in multiple systems \citep{2010ApJS..190....1R}. It also depends on the environment where these stars are formed, finding a higher multiplicity fraction in stellar clusters or star-forming regions compared to field stars \citep{2018MNRAS.478.1825D}. Thus, understanding the observed properties of binary (or multiple) systems provides a better insight into star-formation processes and the evolution of stars and stellar systems \citep[see][and references therein for a recent review]{2024arXiv240312146E}.

There are several methods to unveil binarity among the current stellar surveys. Using spectroscopic data from APOGEE \citep{2017AJ....154...94M}, \citet{2020ApJ...895....2P} characterised the shift in the absorption lines due to the motion generated by the star's orbit in a binary system. \citet{2021MNRAS.506.2269E} used \Gaia EDR3 \citep{2021A&A...649A...1G} astrometry to generate a catalogue of wide binary systems, using the common proper motions and parallaxes of individually resolved components. \citet{2024MNRAS.527.8718W} used a combination of photometric data from \Gaia \citep{2016A&A...595A...1G}, 2MASS \citep{2mass} and WISE \citep{2010AJ....140.1868W} to detect the excess of light in a H-R diagram produced by the presence of a companion. However, most binary systems will remain unresolved and undetected given their intrinsic properties, distance and the limitations of our observations.  

The third \Gaia data release \citep[\Gaia DR3,][]{2023A&A...674A...1G} contains around $1.5 \cdot 10^9$ sources with a full, single-star astrometric solution, but potentially contains a large number of binary systems. \citet{2020MNRAS.495..321P} showed how the presence of an unresolved companion may affect the single-star solution provided by a survey such as \Gaia. Using this effect, \citet{2020MNRAS.496.1922B} demonstrated how the binary properties will impact the \Gaia DR2 \citep{2018A&A...616A...1G} astrometric solution and how this may be indicated by an elevated value of the Renormalized Unit Weight Error (\ruwe). With \ruwe being an effective indicator to pinpoint stellar binarity, different \ruwe thresholds to indicate good single-star solutions, thus flagging "bad" astrometric solutions potentially caused by unresolved binaries, have been proposed for \Gaia DR2 \citep[$1.4$,][]{LL:LL-124} and \Gaia EDR3 \citep[$1.25$,][]{2022MNRAS.513.5270P}. In this paper, we go a step further, and we i) estimate the maximum \ruwe value compatible with single stars as a function of the position on-sky, and ii) characterise what population of binary systems is accessible through this cut on \ruwe.

This paper falls within the scope of the GaiaUnlimited project\footnote{\url{https://gaia-unlimited.org/}}, which aims to provide selection functions and tools for their practical usage, for the \Gaia catalogue and its different data products. This paper thus complements the work done in the estimation of selection functions for the \Gaia DR3 catalogue \citep{2023A&A...669A..55C}, for any subsample drawn from the main catalogue \citep{2023A&A...677A..37C} and for the combination of \Gaia and APOGEE \citep{2024arXiv240105023C}. Similarly to the previous works, the code and tools developed for this paper will be provided through the GaiaUnlimited Python package\footnote{\url{https://github.com/gaia-unlimited/gaiaunlimited}}.

This paper is organised as follows. Section~\ref{sec:Gaia_obs} describes how \Gaia observes unresolved binary systems and the details of the calculation of the key metric (\ruwe) to identify binarity. In Sect.~\ref{sec:binarydetectability}, we explore what properties of binary systems will enhance their detectability in \Gaia. Section~\ref{sec:gums} explores the properties of the unresolved binary systems that may be present in the \Gaia Catalogue of Nearby Stars \citep[GCNS,][]{2021A&A...649A...6G}. Finally, the summary of the paper and our conclusions are presented in Sect.~\ref{sec:conclusions}. We show some examples of how to use the GaiaUnlimited python package in the Appendix~\ref{sec:gaiaunlimited}. 

\section{Observation of unresolved binary systems}
\label{sec:Gaia_obs}

\Gaia has continuously scanned the sky since 2014, registering the times when each source crosses the focal plane \citep{2016A&A...595A...1G}. Subsequently, all these observations undergo processing via the Astrometric Global Iterative Solution \citep[AGIS,][]{2021A&A...649A...4L,2021A&A...649A...2L} which generates the measurements published in the catalogue. AGIS constructs an astrometric track representing a single source when a source has accumulated five valid detections. This process yields fitted astrometric parameters alongside metrics for assessing the quality of the fit, especially \ruwe which we rely on in this present work \citep{LL:LL-124}, among other criteria. In essence, \ruwe serves as a metric to distinguish between good and bad AGIS solutions using a singular value. Typically, it approximates unity for well-behaved individual sources, while higher values indicate a poorer fit by AGIS.

\begin{figure}
    \centering
    \includegraphics[width = 1.\columnwidth]{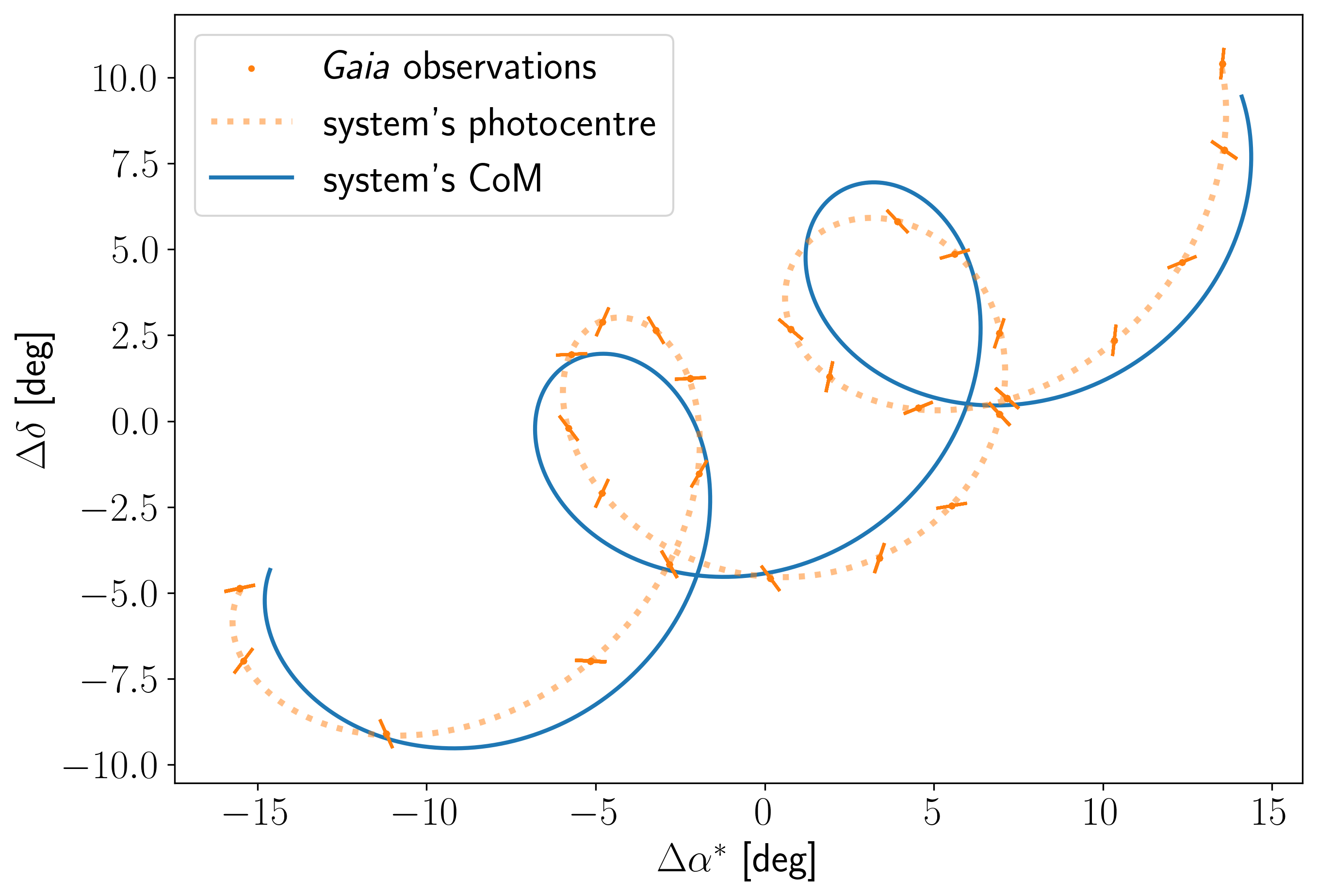}
    \caption{Astrometric track of a binary system seen by \Gaia. The solid blue line tracks the motion of the CoM of the system, while the dotted orange line tracks its photocentre. \Gaia individual scans of the sources are marked with orange solid lines aligned with the observation scanning angle. The goodness-of-fit of the CoM to the \Gaia observations for this simulated system result in a \ruwe of $2.07$.}
    \label{fig:Gaia_track}
\end{figure}

There are several causes for a bad AGIS fit (see Sect.~\ref{sec:calibration}), evident in a \ruwe value significantly exceeding one, which often arise from the wobbling of a source's photocentre during the \Gaia observations. This is the case for binary systems (or multiple stellar systems, in general), where the centre of mass (CoM) and the photocentre of the system follow distinct trajectories depending on the system's characteristics. This effect is shown in Fig.~\ref{fig:Gaia_track}, where the photocentre of a binary system (orange dotted line) oscillates around its CoM (indicated by the blue line), while the CoM itself traces the trajectory fitted by AGIS for a single source. The mismatch between the CoM and the photocentre results in an estimated \ruwe of $2.07$ for this system. Therefore, given that the \ruwe effectively captures the presence of an unresolved companion, it serves as a valuable indicator of binarity in \Gaia sources \citep{2020MNRAS.496.1922B,2021ApJ...907L..33S,2022A&A...657A...7K}. 

\subsection{\ruwe for binary systems}
\label{sec:ruwe}

Estimating the \ruwe for unresolved binary systems involves computing the separation between the photocentre and the CoM in the along-scan direction (AL), which will depend on both the \Gaia observations and the configuration of the binary system, while considering the astrometric error ($\sigma_{\text{ast}}$) which depends on the \Gaia telescope configuration. As described in \citet{LL:LL-136} \citep[see also Sect.~4.1 in][]{2022MNRAS.513.2437P}, the equations governing a five-parameter solution for noisy observations are described by
\begin{align}
    &\Delta\alpha^* \, \sin\psi \, + \Delta\delta \, \cos\psi \, +  \nonumber\\
    &\Delta\varpi \, P_{AL} \, +  \nonumber\\
    &\Delta\mu_{\alpha^*} \, \tau \, \sin\psi \, + 
    \Delta\mu_\delta \, \tau \, \cos\psi \, = \delta\eta + \mathcal{N}(0,\sigma_{\text{ast}}),
\label{eqn:obs_bias}
\end{align}
where $\psi$ represents the scanning angle, $\tau = t - t_{\text{ref}}$ where $t$ is the observation time in years and $t_{\text{ref}} = 2016.0$, $P_{AL}$ stands for the parallax factor, $\delta\eta$ is the AL bias (derived from the projected AL separation between the CoM and the photocentre) and $\sigma_{\text{ast}}$ is the AL astrometric error. This system comprises $n_\text{obs}$ equations, where $n_\text{obs}$ is the number of CCD observations \citep[source transits on \Gaia's focal plane typically generate $9$ CCD observations][]{2016A&A...595A...1G}, and is solved for $\Delta\alpha^*$, $\Delta\delta$, $\Delta\varpi$, $\Delta\mu_{\alpha^*}$ and $\Delta\mu_\delta$, which denote the expected biases arising from binarity (and will be zero for single-sources). From the residuals of these solutions, which quantify how much of the binary motion differs from the fitted single source astrometric solution, we can compute the $\chi^2$ statistic, related to \ruwe as
\begin{equation}
    \text{\ruwe} = \sqrt{\frac{\chi^2}{n_\text{obs} - 5}},
\end{equation}
note that the \ruwe value reported in \Gaia DR3 is a rescaled version of the original $\chi^2$ from AGIS \citep[see][for further details]{LL:LL-124}.

The details from the \Gaia scanning law necessary to estimate a \ruwe of a particular system, that is solving Eq.~\ref{eqn:obs_bias}, are accessible through the \Gaia Observation Schedule Tool \citep[GOST\footnote{\url{https://gaia.esac.esa.int/gost/}},][]{GOST}. GOST provides observation times, parallax factors, and scanning angles linked to a given set of sky coordinates. In our case, these coordinates correspond to the pixel centres at the HEALPix level 5 resolution observed between July 25, 2014, and May 28, 2017, covering the 34-month duration of \Gaia DR3 observations. We choose the resolution at HEALPix level 5 to limit the computational cost of the calculations, however, the procedure described here and the examples shown in Appendix~\ref{sec:gaiaunlimited} are general for any spatial resolution. On the other hand, concerning the observed system, we need to compute the projected AL separation between the system's CoM and its photocentre at each observation time. This value is either zero (applicable for single sources, as discussed in Sect.~\ref{sec:threshold}) or computed based on the binary system configuration (as detailed in Sect.~\ref{sec:binarydetectability}). Lastly, the AL astrometric error $\sigma_{\text{ast}}$ is determined as a function of the G magnitude, following the relationship provided in \citet[][see their Fig.~A.1]{2021A&A...649A...2L}. To obtain its specific numerical value, we use the function \texttt{sigma\_ast} available within the \texttt{astromet} python package\footnote{\url{https://github.com/zpenoyre/astromet.py}}.

\subsection{\ruwe threshold for single stars}
\label{sec:threshold}

Section~\ref{sec:ruwe} describes the recipe to forward model a \ruwe value for a binary system based on its properties and location on-sky. However, to flag these binary systems we need to understand what \ruwe values would describe a well-behaved single-source solution. For this, we characterise the maximum allowed \ruwe, or \ruwe threshold, which will be compatible with single sources as a function of the sky coordinates. Following our approach, we estimate \ruwe values for a population of $10^5$ simulated single sources located at the centres of each HEALPix at level 5 and observed under \Gaia DR3. In this case, where the source’s CoM and photocentre follow identical tracks ($\delta\eta = 0$), the expected \ruwe value will be a $\chi$ distribution scaled by $1/\sqrt{(n-5)}$, which for large $n$ will approach a Gaussian with a dispersion that is proportional to the AL astrometric error. We assign a magnitude to each simulated single source based on a uniform distribution within $G \in (5,20.7)$ mag, to cover the entire magnitude range. The corresponding $\sigma_{\text{ast}}$ value is computed based on this magnitude. Similar to the approach by \citet[][see their Fig.~4]{2022MNRAS.513.2437P}, we fit a Student's t-distribution to the resulting distribution of \ruwe. From this distribution, we define the maximum \ruwe value compatible with the star being a single source, allowing for a one-in-a-million confusion rate, i.e., solving for \ruwe where \mbox{$1 -$ P(\ruwe\,<\,$\tau_\text{\ruwe}$) $= 10^{-6}$}, where $\tau_\text{\ruwe}$ is the computed \ruwe threshold. This process is repeated $30$ times to account for the random errors and to estimate an associated uncertainty ($\leq 2\cdot10^{-3}$) for this \ruwe value. The final simulation-based \ruwe threshold is the mean value of these $30$ realisations.

\subsubsection{Calibration with \Gaia data}
\label{sec:calibration}

So far, we focused on the methodology to calculate the \ruwe value for a \Gaia observation of an isolated source, considering the presence or absence of an unseen companion. This methodology is particularly useful for modelling \ruwe values for unresolved binary systems and deriving general conclusions for this population (see Sect.~\ref{sec:binarydetectability}). However, other sources' properties such as stellar crowding, variability or the presence of disc material around single stars \citep[among others,][]{2020MNRAS.496.1922B,2022RNAAS...6...18F}, or attitude and geometric calibration errors and other instrumental properties, also contribute to the increase in the sources' \ruwe \citep{LL:LL-124}. In regions with particularly high stellar density, the observed single source might be influenced by neighbouring stars causing \Gaia's measured image to deviate from the single source PSF model. This deviation leads to difficulties in accurately determining the PSF's centroid, resulting in an elevated \ruwe. 

\begin{figure}
    \centering
    \includegraphics[width = 1.\columnwidth]{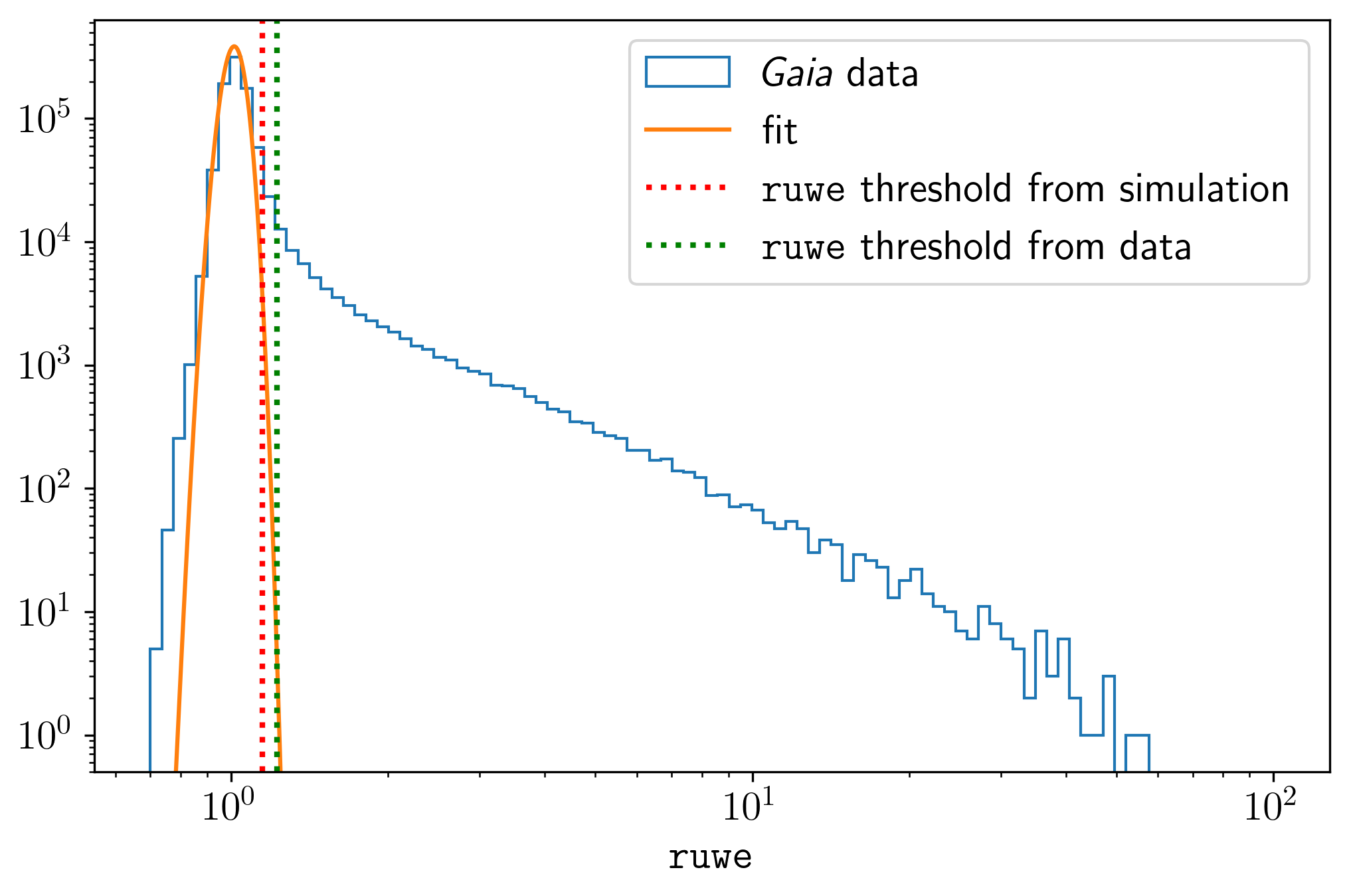}
    \caption{Histogram of \ruwe for the sources in the \Gaia archive (blue line) HEALPix centred at $l = 73.58^\circ$, $b = 3.17^\circ$. The solid orange line indicates our fit of the single source population in that HEALPix. The dotted lines indicate the values of the \ruwe threshold for single sources with (green) and without (red) crowding effects.}
    \label{fig:ruwe_fitData}
\end{figure}

A detailed modelling of the previous effects on the source's \ruwe is out of the scope of this paper. Instead, we gauge these effects, specifically focusing on crowding, directly from the \Gaia data employing the following approach. For HEALPix at level 5, we use \texttt{scipy}'s \texttt{curve\_fit} function in Python\footnote{\url{https://scipy.org/}} to fit a Gaussian to the left half of the \ruwe distribution of all the sources in the HEALPix retrieved from \Gaia archive (left half with respect to the mode of the distribution, usually centred at one but automatically estimated from the data). Using the full Gaussian, representing a single-source population, we obtain the maximum \ruwe allowed for single sources by applying the procedure described in Sect.~\ref{sec:threshold}. Figure~\ref{fig:ruwe_fitData} shows the \ruwe distribution for the $\sim 10^6$ sources in \Gaia corresponding to the HEALPix centred at $l = 73.58^\circ$ and $b = 3.17^\circ$. Additionally, we show the Gaussian distribution fitted to the sources centred around \ruwe $\sim 1$. In this case, the \ruwe threshold for single sources estimated from both the simulated single-source population (see Sect.~\ref{sec:threshold}) and actual data are $1.14$ and $1.22$, respectively. 

The data-driven approach adopted here faces limitations in its application across the whole sky due to the necessity of a minimum source count, and the \ruwe distribution around its mode to be well represented by a Gaussian for an accurate fit. For HEALPix regions where we obtain a data-driven \ruwe threshold, we fit a second-order polynomial to the difference of the \ruwe thresholds, i.e., $\tau_{\ruwe_\text{data}}\,-\,\tau_{\ruwe_\text{simulation}}$, as a function of the logarithm of the source count within that HEALPix. With this, we derive an adjustment factor, or offset, to refine the simulation-based \ruwe threshold accounting primarily for the influence of crowding based on the number of sources present within each HEALPix.

\begin{figure*}
    \centering
    \includegraphics[width = 1.\textwidth]{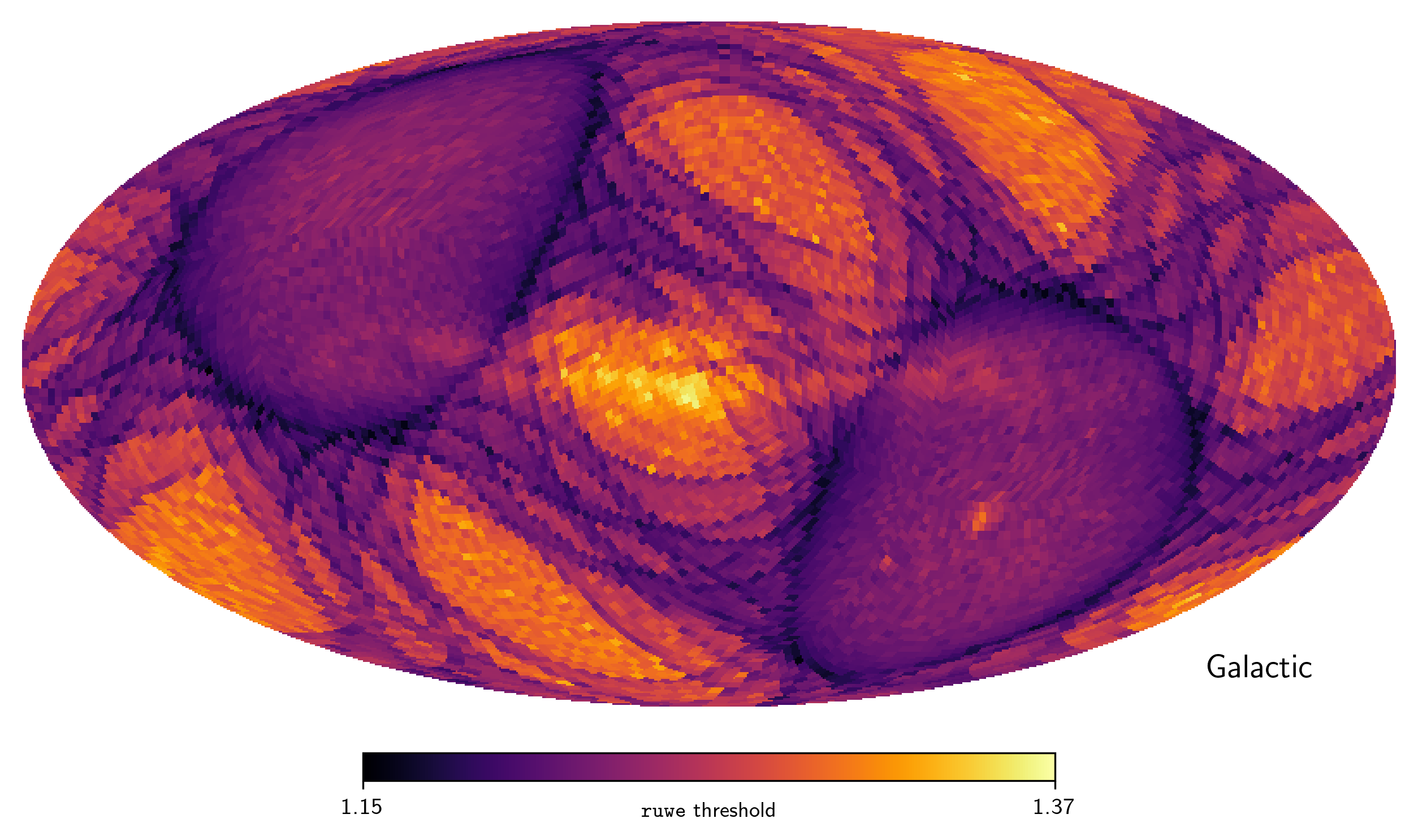}
    \caption{Sky map of the sky-varying \ruwe threshold at the resolution of HEALPix level 5. Sources above these \ruwe threshold indicate potential binary systems. These \ruwe values show a strong influence of the \Gaia scanning law, while also highlighting regions with high source density.}
    \label{fig:ruwe_singleSource}
\end{figure*}

Figure~\ref{fig:ruwe_singleSource} shows the estimated upper value of \ruwe compatible with single star solutions across the sky. It shows a strong dependence on the \Gaia scanning law, indicating that regions with a higher number of observations tend to exhibit lower \ruwe thresholds since the fitted AGIS solution for each source is better constrained. We also see the impact of crowding within the Galactic plane, especially toward the Galactic centre and the Magellanic clouds. These regions show elevated \ruwe thresholds due to challenges in reliably fitting the single source track using AGIS caused by the misidentification of sources due to their high stellar density. This misidentification for a large fraction of sources in those regions will cause the single source \ruwe distribution to be skewed towards \ruwe larger than one, resulting in a higher threshold value after our calibration.

The \ruwe threshold variation across different sky coordinates presented in our work represents an improvement compared to prior studies that employed a singular \ruwe value across the whole sky. For instance, in \Gaia DR2, \citet{LL:LL-124} suggested a \ruwe threshold of $1.4$ to characterise good \Gaia solutions, with higher values potentially indicating binarity or other influencing factors. For \Gaia EDR3, \citet{2022MNRAS.513.2437P} investigated the imprints left in the \ruwe distribution attributed to nearby binary systems (within $100$ pc) and recommended a threshold of $1.25$ for identifying potential binarity (the authors' discussion focuses around \texttt{luwe}, which stands for local-renormalised \texttt{uwe}). These different \ruwe thresholds suggested in different studies highlight the sensitivity of \ruwe to different observational factors, their relation to specific \Gaia data releases, e.g., DR2 or DR3, and the characteristics of the observed regions. 

\subsubsection{Selection probability of sources with high \ruwe}
\label{sec:binarySF}

We compute the selection probability for sources in \Gaia DR3 with a value for \ruwe higher than the estimated \ruwe threshold, using the approach described in \citet{2023A&A...677A..37C} to estimate selection functions for \Gaia subsamples. We note here that for a universe consisting only of single stars without any photocentre motion, and for perfectly behaved and calibrated \Gaia instruments, by construction the probability to select sources above the \ruwe threshold would be $10^{-6}$, independent of sky position or source brightness. In reality (as shown below) the selection probability is much higher and this can indicate, among others, the presence of unseen companions to the selected sources.

\begin{figure}
    \centering
    \includegraphics[width = 1.\columnwidth]{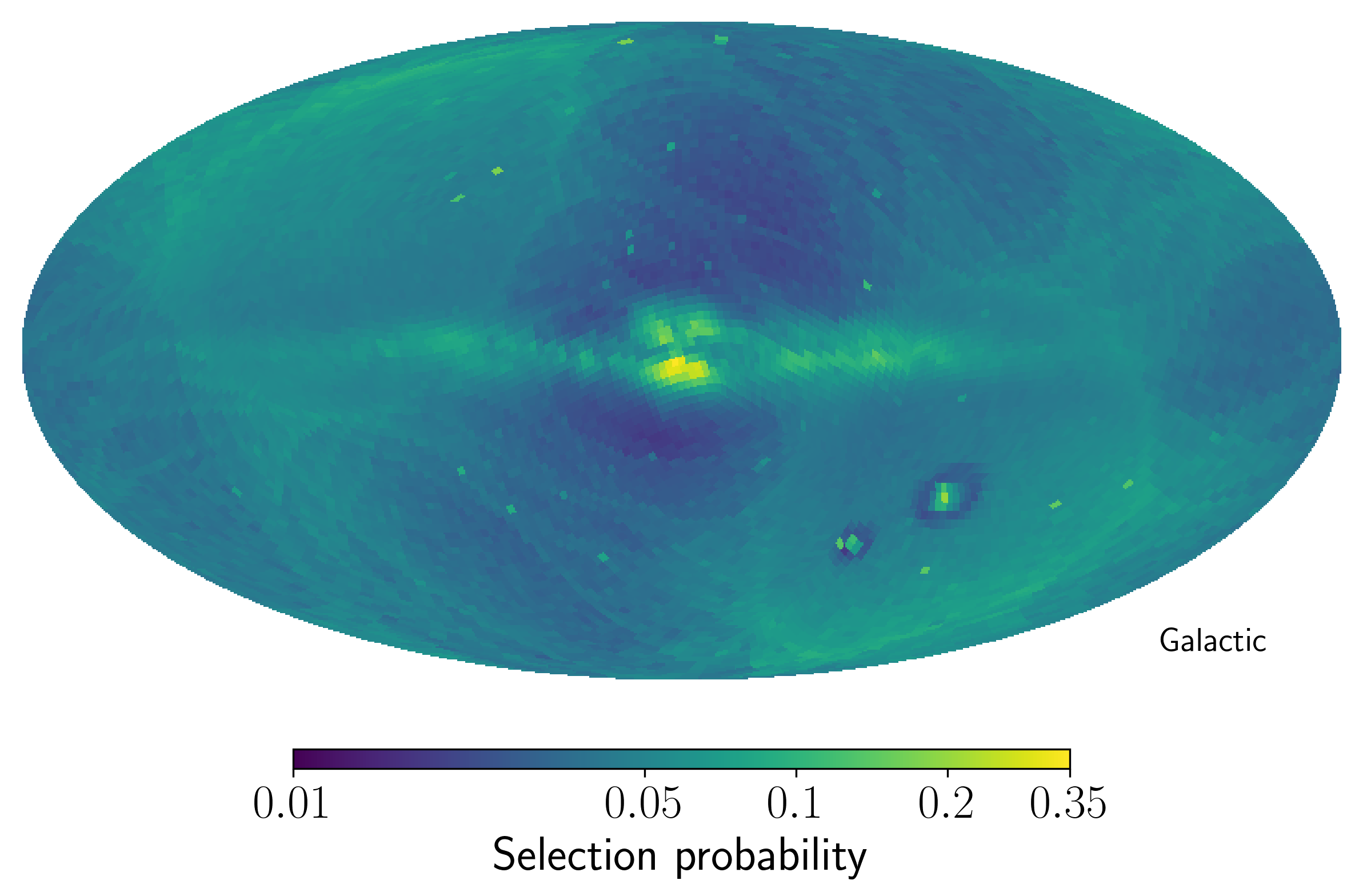}
    \caption{Selection probability for sources with a \ruwe higher than the threshold represented in Fig.~\ref{fig:ruwe_singleSource}. The regions with higher selection probability, e.g., cores of globular clusters, the Galactic plane and the Magellanic Clouds, contain sources with a \ruwe increased due to a combination of binarity, source crowding and other effects. These regions may indicate a higher contamination rate when selecting binaries solely based on \ruwe. The dark halos around the Magellanic Clouds are due to an excess of faint sources with low \ruwe in that region.}
    \label{fig:binaries_SF}
\end{figure}

Within each HEALPix at level 5, we determine the expected selection probability as 
\begin{equation}
E(p) = \frac{k+1}{n+2},
\label{eqn:expected_value}
\end{equation}
where $k$ represents the count of sources exceeding the estimated sky-varying \ruwe threshold (shown Fig.~\ref{fig:ruwe_singleSource}) and $n$ is the total number of sources within that HEALPix. Figure~\ref{fig:binaries_SF} shows the estimated selection probability across the sky, revealing that the selection of these sources, which in some regions may translate to the detectability of binary systems defined as sources with large \ruwe, correlates with the \Gaia scanning law. We see a higher selection probability in crowded areas, e.g., the Galactic plane, particularly towards the Galactic centre, and the Magellanic Clouds or the hot pixels corresponding to the cores of globular clusters. We can't interpret these high-selection probability regions as being highly populated by binary systems. As previously discussed, several reasons in both the sources or instrument properties can contribute to an increased \ruwe value. Therefore, these high-selection probability regions may be seen as regions with an important contribution of factors other than binarity to increase the sources' \ruwe, tracing regions of high contamination when binary systems are selected solely based on the latter. In this paper, our modelling primarily focuses on the increase in \ruwe caused by a bad single-source fit by AGIS due to the presence of an unresolved companion, which is typically the predominant contribution across most of the sky, i.e. Galactic latitudes above the plane and avoiding crowded regions. Consequently, selecting binary systems based on \ruwe within highly crowded areas will include a higher level of contamination or interference from other contributing factors, even with our sky-varying \ruwe threshold taking crowding into account.
 
\section{Properties of detected binary systems}
\label{sec:binarydetectability}

The approach presented in Sect.~\ref{sec:Gaia_obs} allows us to estimate \ruwe values for any source observed by \Gaia, including unresolved binary systems, based on their properties and location on-sky. With this approach, and considering a specific population of binary systems, we can characterise what properties will enhance the probability of detection for unresolved binary systems in \Gaia~DR3, when detected as systems whose estimated \ruwe is higher than the threshold shown in Fig.~\ref{fig:ruwe_singleSource}.

\subsection{Simulated binary population}
\label{sec:binarypopulation}

We simulate a population of binary systems to predict their detectability as a function of the system's parameters. This simulated population is not intended to mirror a realistic representation of binary systems distributed throughout the Galaxy. Instead, the simulation aims to cover a broad range of stellar binary characteristics, e.g. mass and light ratios or orbital properties, general enough to allow for an exploration of the effect of the binary parameters on the probability of being selected based on a \ruwe threshold which varies across the sky.

We generate around one million MS binaries to study the detectability variations of different simulated parameters. First, we retrieve the stellar parameters from the PARSEC\footnote{\url{http://stev.oapd.inaf.it/cgi-bin/cmd}} isochrones \citep{2012MNRAS.427..127B} for a uniform grid of different ages. We isolate the isochrones' MS and take the corresponding masses as the mass of the binary primary component. We assign a mass ratio ($q$) following a uniform distribution from $0$ to $1$ to each of the primary masses to build a binary system. These systems are uniformly placed at varying distances ranging from $10$pc to $5$kpc. We calculate \Gaia's $G$ magnitude from the mass and distance distributions, considering no extinction, and restrict our analysis to systems within \Gaia's observational limits, i.e., $G < 20.7$ mag. The remaining parameters for the simulated binaries follow the distributions described in Table~\ref{tab:binary_parameters} and are similar to those used in \citet[][their Table~1]{2022MNRAS.513.2437P}. The semi-major axis of the binary orbit is computed using the masses of both components and the orbital period. Finally, since we consider MS-MS binaries, the luminosity ratio is computed from the mass ratio as $l = q^{3.5}$.

\begin{table}
    \centering
    \renewcommand{\arraystretch}{1.2}
    \begin{tabular}{lr}
         Parameter& Distribution\\
         \hline
         \hline
    Distance $[\text{pc}]$           & $U(10,5000)$\\
    $\log\,(\,$Period $[\text{yr}]\,)$        & $U(-2.56,2)$\\
    Mass ratio                           & $U(0,1)$\\
    Eccentricity                         & $U(0,1)$\\
    Initial phase $[\text{rad}]$         & $U(0,2\pi)$\\
    Azimuthal view angle $(\phi)$ $[\text{rad}]$         & $U(0,2\pi)$\\
    $\cos\,(\,$Polar view angle $(\theta)$ $[\text{rad}]\,)$ & $U(-1,1)$\\
    Planar projection angle $(\omega)$ $[\text{rad}]$         & $U(0,2\pi)$\\
    \end{tabular}
    \caption{Distributions for the simulated binary system population.}
    \label{tab:binary_parameters}
\end{table}

\subsection{Selection probability and binary system properties}
\label{sec:detection_binary}

\begin{figure}
    \centering
    \includegraphics[width = 1.\columnwidth]{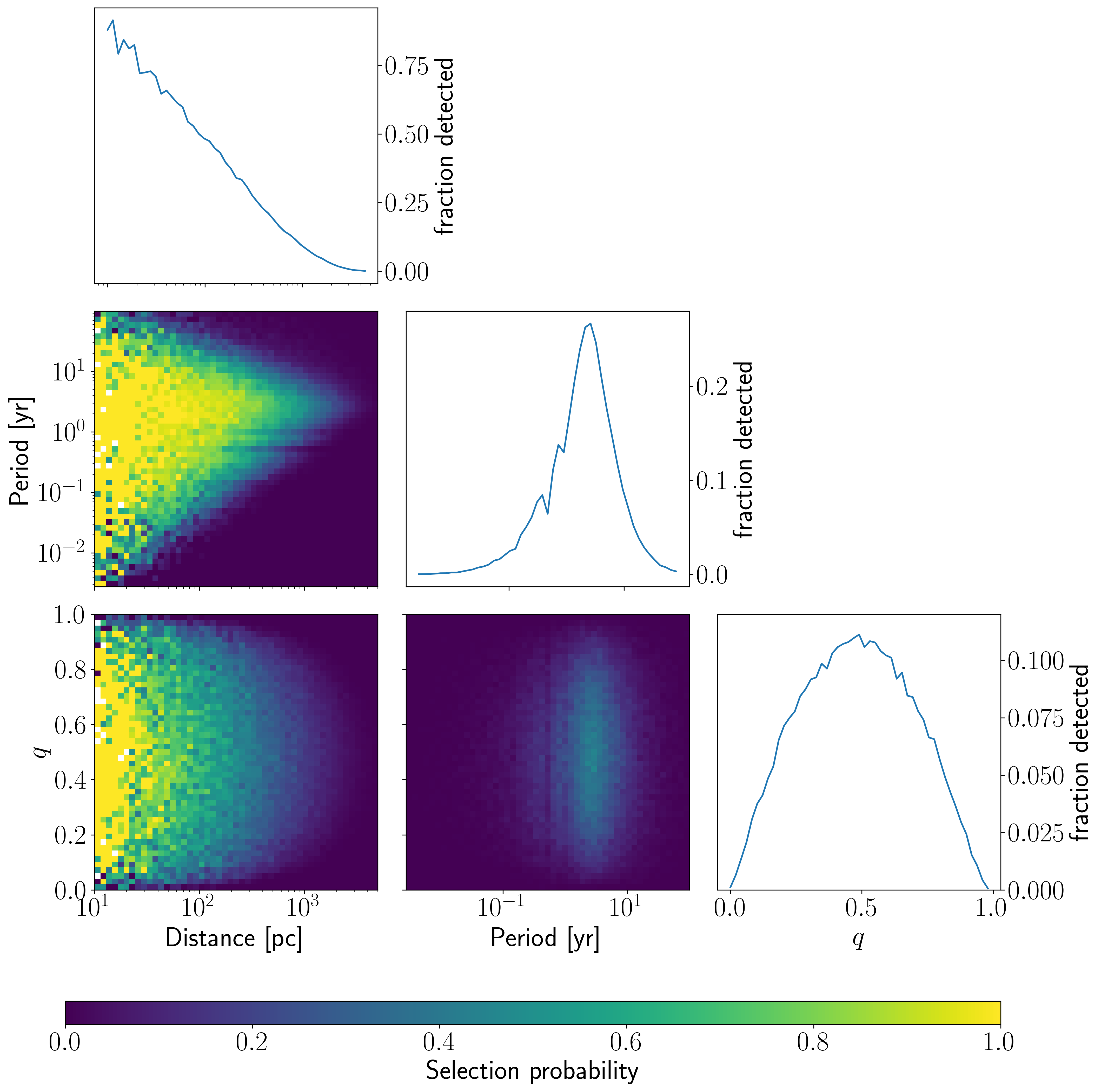}
    \caption{Heatmap of binary systems' properties, i.e., distance, period and mass rations, that will enhance their detection. The detectability of these systems monotonically decreases with distance, reaching a detection fraction of $50\%$ at around $100$ pc. There is a strong detectability window around periods of three years, corresponding to the \Gaia observation time baseline. Systems with mass ratios of zero, i.e., single sources, or one show no sign of binarity on \ruwe since there is no wobbling of their CoM around the photocentre.}
    \label{fig:detectability_cornerPlot}
\end{figure}

We distribute the generated population of binary systems across various sky locations, to evaluate the fraction of these systems that would be detected using \ruwe (estimated as in Sect.~\ref{sec:ruwe}). Figure~\ref{fig:detectability_cornerPlot} shows this detection fraction for the different binary configurations considered at the pixel centred at $l = 15.3^\circ$ and $b = -14.4^\circ$, with yellow pixels representing a nearly complete detection while bluer pixels indicating a detection fraction closer to zero within that specific configuration. 

As expected, the binary detectability peaks at a mass ratio of $q = 0.5$. No systems are detected when $q = 0$ (equivalent to single sources), or when $q = 1$ where the CoM and the photocentre trace the same trajectory, which means the value of \ruwe is not changed.

The detection probability shows a peak at an orbital period aligning with the observational baseline of the survey, which is 34 months for \Gaia DR3. This probability gradually declines for longer periods, given that no observable orbital motion is evident within this observation time \citep{2020MNRAS.495..321P}. Furthermore, we observe notable declines in the fraction of detected systems when the orbital period coincides harmonically with a year, prominently visible in Fig.~\ref{fig:detectability_cornerPlot} at periods of 1 year and 1/2 year. This decrease in detectability arises due to a confusion of the orbital motion with a parallax shift, leading to a situation where a seemingly good astrometric fit (resulting in low \ruwe) corresponds to an erroneous parallax. These features in the period distribution are also visible in the astrometric binaries in \Gaia DR3 \citep[][also selected based on \ruwe]{2023A&A...674A...9H}. We observed that the strength of this effect, however, depends on the direction of the observed binary systems (details of the \Gaia scanning law).

\begin{figure}
    \centering
    \includegraphics[width = 1.\columnwidth]{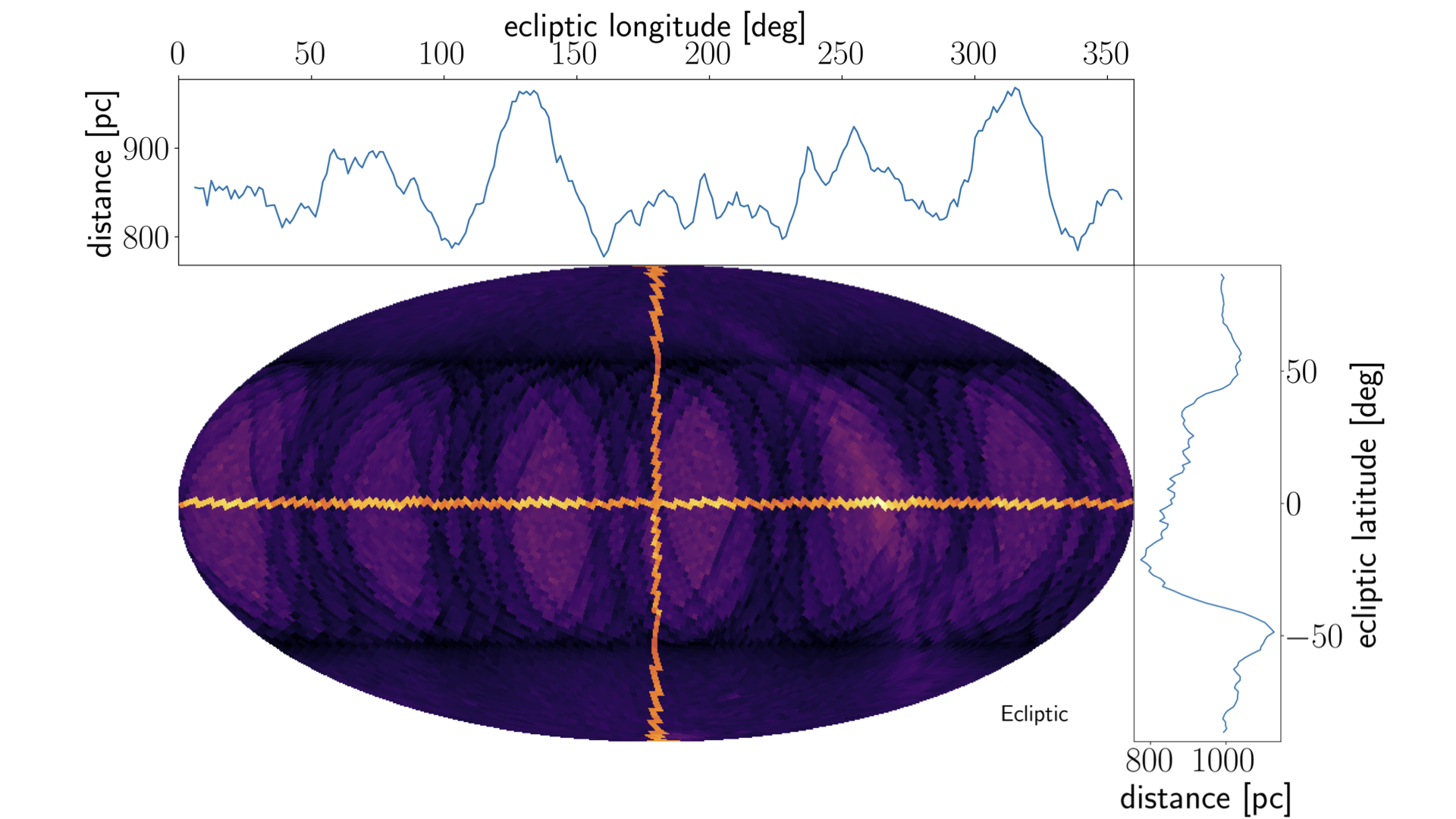}
    \caption{Central panel: Same as in Fig.~\ref{fig:ruwe_singleSource}, in ecliptic coordinates. The pixels highlighted in the vertical and horizontal directions show the regions where we estimate the maximum distance where we detect half of the simulated systems, which is shown in the right and top panels, respectively. Top panel: Distance where $50\%$ of the systems are detected, for the highlighted pixels along the ecliptic $\lambda = 0^\circ$. Right panel: Same as the top panel, for ecliptic $\beta = 0^\circ$.}
    \label{fig:binaries_maxDist}
\end{figure}

From the first column in Fig.~\ref{fig:detectability_cornerPlot}, we see that our ability to detect binary systems is monotonically decreasing with distance, reaching $50\%$ of systems detected at around $100$pc. To capture the variation of this behaviour across different regions of the sky, we estimate the \ruwe of the simulated binary population described in Sect.~\ref{sec:binarypopulation} in different pixels from the north to the south ecliptic poles at ecliptic longitude $\lambda = 0^\circ$, and from $\lambda = 0^\circ$ to $\lambda = 360^\circ$ at ecliptic latitude $\beta = 0^\circ$, as shown in Fig.~\ref{fig:binaries_maxDist}. We use ecliptic coordinates since the \Gaia scanning law is symmetric in this coordinate system, and the strongest contribution to the variation of the \ruwe threshold across the sky (see Sect.~\ref{sec:threshold}). Also, we focus on periods from one to four years which is the window for optimal detectability, since most of the selected binary systems will be in this period range (see discussion in Sect.~\ref{sec:periodWindow}). Figure~\ref{fig:binaries_maxDist} shows how the variation of the distance where $50\%$ of the simulated systems are detected (where the median \ruwe is equal to the \ruwe threshold shown in Fig.~\ref{fig:ruwe_singleSource}). We see that in the ecliptic poles, and up to $|\beta| = 45^\circ$, where \Gaia has uniformly scanned the sky with a large number of observations, we can detect binary systems at larger distances (up to around 1 kpc) due to a more constrained astrometric solution, thus resulting in a lower \ruwe. For the $|\beta| < 45^\circ$, on the other hand, the distance where we detect at least $50\%$ of the simulated binary systems can decrease down to $800$pc. Along the $\lambda$ direction, the distance reached shows a more irregular behaviour since this region captures larger variations in the \Gaia scanning law. However, the distances where we can detect systems can vary $\sim 200$pc depending on the scanning configurations. If we include in the analysis the whole range of simulated periods, the distances where we detect $50\%$ of the simulated systems range from around $60$ up to $150$pc, approximately, depending on the $(\lambda,\beta)$ direction.
 
\section{The GCNS binary population through GUMS}
\label{sec:gums}

\begin{figure}
    \centering
    \includegraphics[width = 1.\columnwidth]{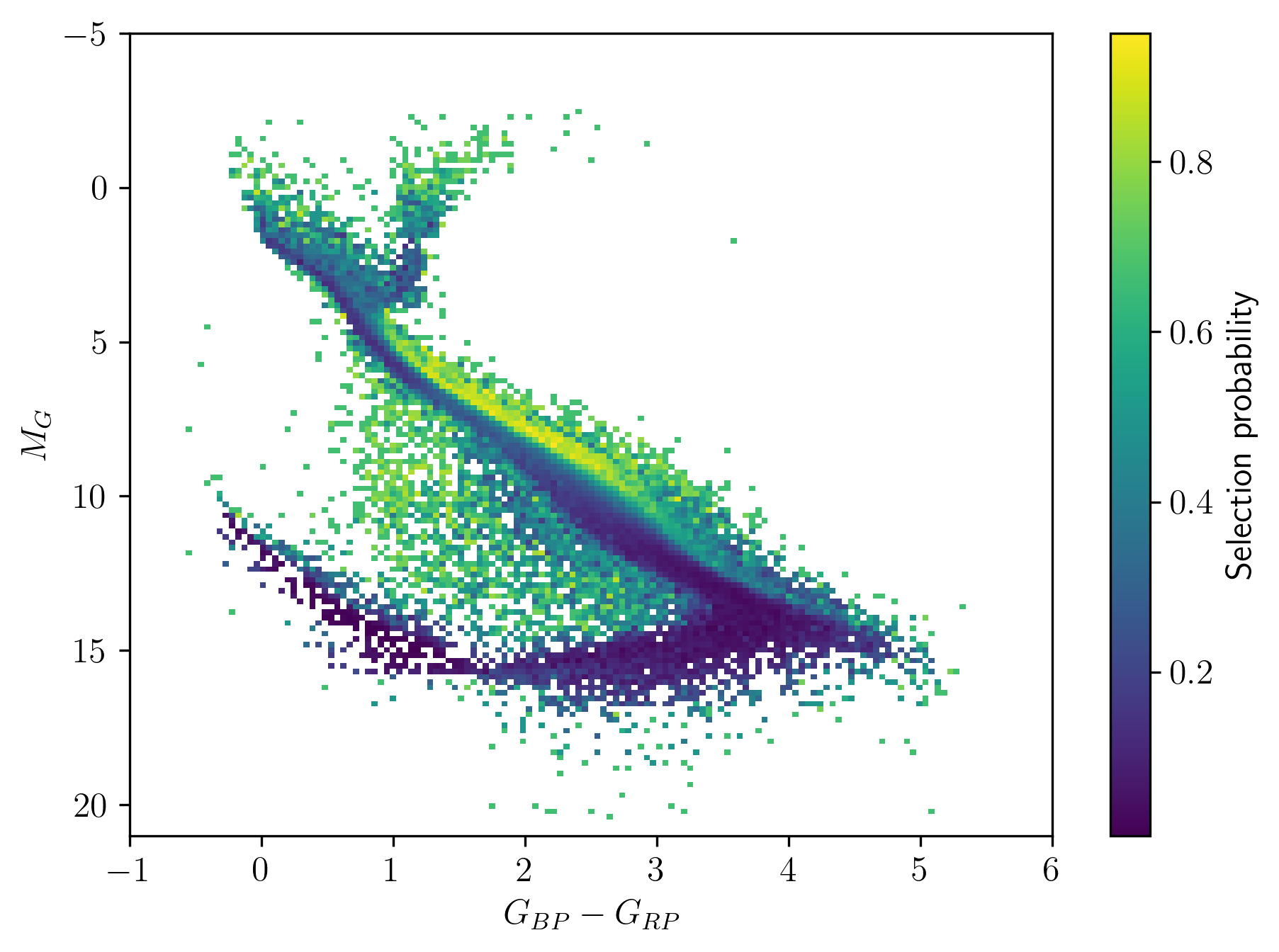}
    \caption{Selection probability of the GCNS sources with \ruwe larger than our threshold shown in Fig.~\ref{fig:ruwe_singleSource} across the colour-magnitude diagram. For the MS and WD sequence, we see a gradient on the selection probability for bins at the same $G_{BP}-G_{RP}$ colour. The region between the MS and the WD sequence, which contains the MS-WD binaries, has a high concentration of sources labelled with poor astrometric solutions in the GCNS}
    \label{fig:gcns_cmd}
\end{figure}

\citet{2021A&A...649A...6G} produced a catalogue of $331\,312$ objects within $100$ pc which was released together with \Gaia EDR3. The catalogue includes a list of probable members to the two closest open clusters, i.e., the Hyades and Coma Berenices, white dwarfs (WDs) and resolved binary systems. It also includes numerous sources with \ruwe significantly higher than one, which may be potential unresolved binary systems that may have gone unnoticed. We find $74\,785$ potential unresolved binary systems with a catalogued \ruwe higher than the threshold for single sources we computed in Sect.~\ref{sec:threshold} (see Fig.~\ref{fig:ruwe_singleSource}). Based on our approach, we can select more binary system candidates compared to the selection based on \ruwe higher than $1.4$ and $1.25$, which provide $55\,754$ and $70\,848$, respectively. In Fig.~\ref{fig:gcns_cmd} we show how our selection of unresolved binary systems is distributed in a colour-magnitude diagram (probability estimated as in Eq.~\ref{eqn:expected_value} with respect to the colour and magnitude bins). We observe a high selection probability in the brighter MS, which monotonically decreases towards faint magnitudes. A similar trend is observed in the WD sequence ($10 \lesssim M_G \lesssim 15$ and $0 \lesssim G_{BP}-G_{RP} \lesssim 1$), where the selection probability is higher in the brighter sequence. The region between the MS and the WD sequence, which shows a selection probability unexpectedly high, corresponds to sources with "poor astrometric solutions" in the GCNS catalogue \citep[see Section~2.1 in][for further details]{2021A&A...649A...6G}. Although this region also contains MS-WD binaries, the \ruwe for the sources in this region is probably also increased for reasons other than binarity. Therefore, this region may represent a highly contaminated region in our analysis.

To understand the completeness and contamination of this sample, and the types of systems that our method selects among the unresolved binary system candidates we use the \Gaia Universe Model Snapshot \citep[GUMS,][]{2012A&A...543A.100R}. GUMS provides, through the \Gaia Object Generator \citep[GOG,][]{gog}, astrometry, photometry and astrophysical parameters of sources as if they were observed by \Gaia. GUMS only provides around $100\,000$ sources within $100$ pc, therefore, to increase the statistical significance of the analysis we increase the volume to analyse to $200$ pc, resulting in $882\,107$ sources of which $243\,377$ are true binary systems.

We forward model a \ruwe value for these GUMS sources following the recipe in Sect.~\ref{sec:ruwe} (see also Appendixy~\ref{sec:gaiaunlimited} for instructions on how to use our GaiaUnlimited python package). Then, we detect these binary systems based on \ruwe as we have done for the GCNS. We can recover $51\,524$ true binary systems with our variable \ruwe approach, which still results in a more complete sample when compared to \ruwe selections higher than $1.4$ and $1.25$, resulting in $43\,398$ and $50\,431$, respectively. The contamination of our sample is one single source selected as a binary system, as expected by the design of our approach (see Sect.~\ref{sec:threshold}). This contamination is zero in the case of \mbox{\ruwe $> 1.4$}, due to the more restrictive criteria, and $20$ in the case of \mbox{\ruwe $> 1.25$}. Therefore, our approach provides a sample of unresolved binary systems with the best compromise between completeness and contamination when compared to other criteria to select this type of object.

\begin{figure*}
    \centering
    \includegraphics[width = 1.\textwidth]{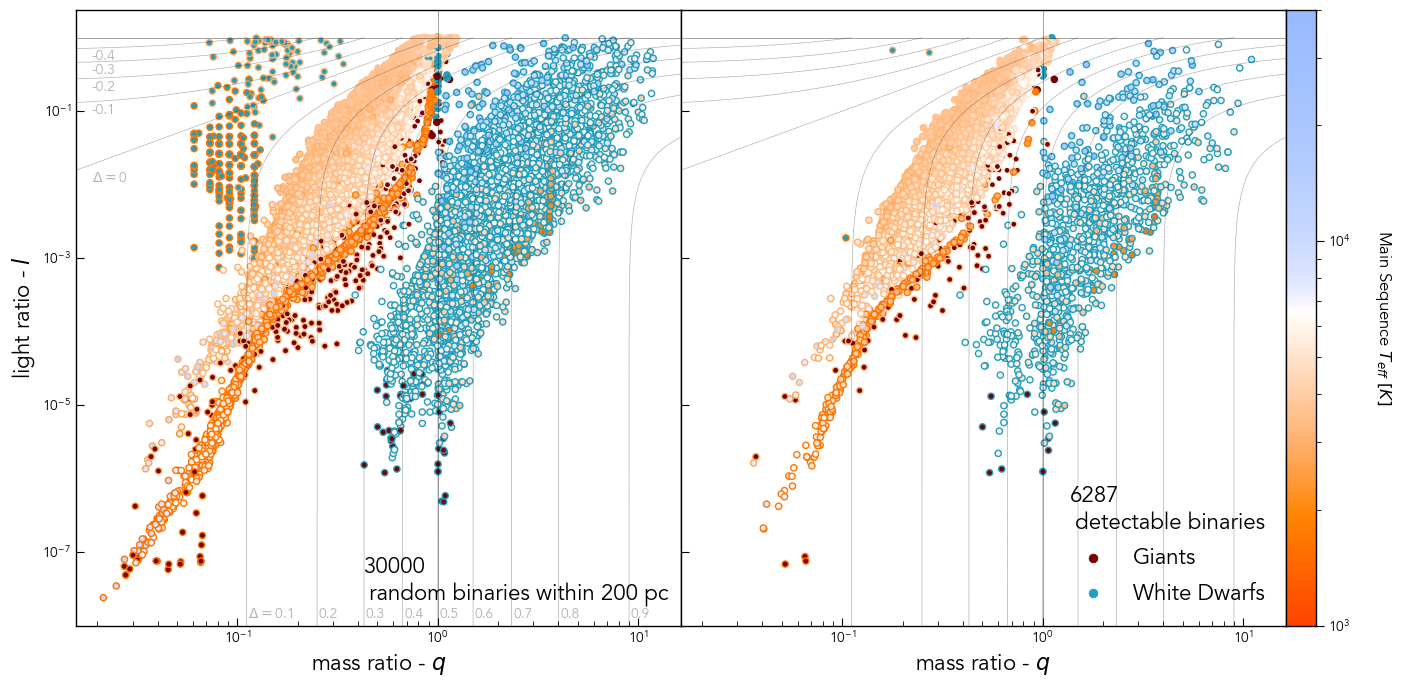}
    \caption{Mass and light ratios of a random subset of 30,000 (of 242,485) binaries from GUMS within 200 pc. The central colour of each point shows the type of the primary, and the outline shows the type of the secondary. MS stars are given a colour that maps approximately to their visual colour based on their effective temperature. Giants ($M_G<4$ and $G_{BP}-G_{RP} >0.9$) are shown in dark red and WDs ($M_G - 3(G_{BP}-G_{RP})<9.5$) in blue. The left panel shows all random binaries and the right panel only those which are unresolved and whose \ruwe is greater than the threshold for their region of the sky. Because we choose the primary to be the brighter component ($l<1$ always) the populations are effectively reflected through $l=1$ and $q \rightarrow \frac{1}{q}$ - thus the population of systems containing WDs appears at both low and high $q$ depending on the relative brightness of the other component. Lines of constant $\Delta =\frac{l-q}{(1+l)(1+q)}$ are shown - describing the relative offset of the centre of light from the CoM (with positive values corresponding to systems where the photocentre is closer to the secondary). }
    \label{fig:qlfinal_rand}
\end{figure*}

Working with simulated data has the advantage that we can access all the true simulated parameters. In this case, we can study the population of binary systems depending on the evolutionary stages of their two components. We have identified five main evolutionary stages present in GUMS, i.e., Giant stars, WD, and high-, low- and mid-MS stars defined as stars with $M_G < 4$, $M_G > 10$ and in-between these cuts, respectively. Figure \ref{fig:qlfinal_rand} shows the distribution of mass and light ratios for GUMS binaries within 200 pc - and the subset which are unresolved and have detectable excess \ruwe. This shows the range of important (though often not observable) physical properties we might expect from a range of binary systems. MS stars, especially those that are cooler and of lower mass, follow a relatively tight distribution (akin to our previous heuristic of $l\propto q^{3.5}$), above this (and thus closer in light ratio for a given mass ratio) we see binaries containing stars the mass and temperature of our Sun and higher, which are significantly less tightly clustered but follow a similar trend. Systems containing Giants generally fall below these populations (as the Giant primary is effectively over-luminous for its mass). Systems containing WDs are quite distinct in this space - having generally much more mass compared to their luminosity (except for WD-WD binaries which cluster very close to $q \approx 1$ and $l \approx 1$). 

The strong dependence of light ratio on mass ratio leads to many unresolved binary systems being indistinguishable from single objects from their luminosity (and also in general from their spectra) as even moderate mass ratios, e.g. $q=0.5$, lead to much smaller light ratios ($l<0.1$). Thus the light ratio, though crucial for fully characterising binary systems, is rarely measurable.

Comparing the full population to the detectable we can get a feeling for which kinds of systems are easier or more difficult to pick out from astrometry alone. We can see a low detection fraction for systems with $\Delta$ close to zero (as the astrometric motion is negligible) - which includes both 'twin' systems ($q\sim l \sim 1$) as well as faint MS-MS and MS-Giant binaries (with $q \lesssim0.1$). The detection fraction is also low for systems containing WDs where the second component is similarly or less luminous (top left of each panel, because these are dim enough that the astrometric error is high enough to dominate over astrometric motion).

\subsection*{Detectability functions with period and distance}
\label{sec:periodWindow}

\begin{figure}
    \centering
    \includegraphics[width = 1.\columnwidth]{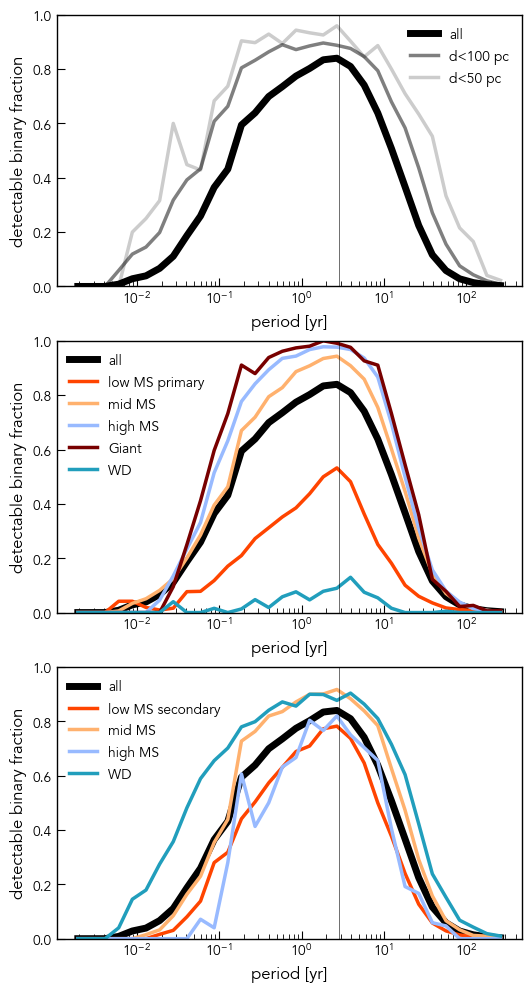}
    \caption{The fraction of detectable binaries as a function of period, comparing the full population (within 200 pc) to subsets. Top panel: detectability within some distance range. Middle panel: detectability dependent on the type of the primary. Bottom panel:  detectability dependent on the type of the secondary. The vertical line at 34 months shows the time baseline of \Gaia DR3 (and is the optimal period for binary detection). MS stars are split into three subgroups - high ($M_G <4$), low ($M_G>10$) and mid (in-between these two cuts). Around 15\% of MS primaries fall into the high bin, and the same for the low. Close to half of the secondaries are low, and the vast majority of the rest are mid (the number of high-mass MS secondaries is small, and Giant secondaries are negligible)}
    \label{fig:period_frac}
\end{figure}

Figure \ref{fig:period_frac} shows the fraction of binaries that are detected as a function of period, expanding on Figure~15 of \citet{2022MNRAS.513.2437P}. This is a very strong 'window function' limiting us to only binaries for which we see a significant fraction of the orbit ($P \lesssim 10$ years, with the fraction dropping as $P^{-2}$) and those whose orbits are wide enough that the astrometric motion is significant compared to the astrometric error \citep[$P \gtrsim 1$ month, with the fraction increasing proportional to $a \propto P^\frac{2}{3}$, see][for more detail]{2020MNRAS.495..321P,2022MNRAS.513.2437P,2022MNRAS.516.3661A}.

Even for a sample out to $200$ pc (which will be dominated by the number of systems close to that distance limit), we detect more than $80 \%$ of systems at the optimal period of 34 months (\Gaia DR3's time-baseline). For a closer limiting distance that fraction increases mostly due to the larger apparent magnitude and parallax, both leading to the astrometric signal being more dominant over astrometric noise.

\begin{figure}
    \centering
    \includegraphics[width = 1.\columnwidth]{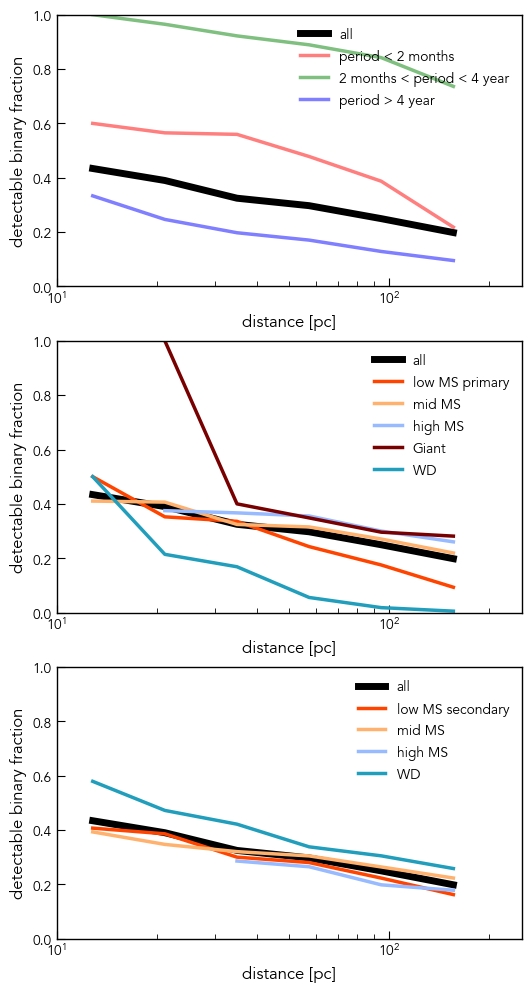}
    \caption{Similar to Fig.~\ref{fig:period_frac} but separated by the distance of the binary. We show the function for all systems in the sample and for subsamples separated by binary period (top), type of primary (middle) and type of secondary (bottom).}
    \label{fig:distance_frac}
\end{figure}

Separating systems by the type of the primary we see essentially the same curve, with brighter primaries more easily detectable (again because of the reduced astrometric noise). Low mass primaries are significantly harder to detect (as the orbit is smaller for the same period) and so are WD primary systems (as these are necessarily low luminosity). The most massive primaries, especially the Giants, are even more strongly peaked in period, particularly at the low period end. This is mostly a general astronomical bias rather than an astrometric one, since these are both rarer (thus they don't appear in significant numbers at small distances and smaller period orbits can be unresolvable at larger distances) and more limited in their minimum period by their large stellar radii and the requirement that the stars do not merge. Comparing instead the detectability based on the type of the secondary we now see that WD companions give significantly larger signals (because of their large $q$ and low $l$, giving large $\Delta$). In general, it is not a bad assumption that each of the curves shown here is a simple rescaling of the distribution across all stars, with the small caveat that it must saturate at a detection fraction of 1. Figure \ref{fig:distance_frac} shows a similar detectability function but now as a function of distance. All populations generally have lower detection fractions at increasing distances, where they are both smaller orbits on-sky and dimmer, thus the astrometric error can increase. 

Evidently, both the period and distance functions apply to the data at once, so we start by showing the detectable distance fraction split by period. For ideal periods (close to the peak at 34 months), the detectable fraction is close to 1 at small distances and still near $80 \%$ at the edge of our sample - as shown in Fig.~\ref{fig:binaries_maxDist} this will only drop below $50 \%$ for sources $\sim$ 1 kpc and further. Longer-period systems are harder to detect but follow the same kind of behaviour, whilst short-period systems drop off non-linearly as the population approaches the astrometric noise floor of \Gaia. Separating by primary type we see that dim primaries (WDs and low-MS stars) also drop off quickly with distance for the same reason, and more massive brighter systems follow the general trend. Finally, separating by secondary type, we again see that WD companions are the easiest to detect, but the slope of the detection fraction with distance does not depend on the secondary type.

For all of these systems, the limiting factor with distance is the ratio of the magnitude of the astrometric motion to the astrometric error, which will necessarily decrease with distance. For populations for which no sources remain in the (approximately) constant regime of astrometric error ($G \gtrsim 13$), this ratio drops more rapidly and thus the curve steepens. We would expect this steepening to apply to all types of sources eventually, modulated by their inherent luminosity, and thus the brightest Giants may still be detectable as binaries at large ($>$ kpc) distances. This estimation does not account for extinction and crowding increasing with distance, both of which will cause extra astrometric noise and may lower the distance at which binary detection is achievable, especially in crowded and obscured regions such as the Galactic disc, Galactic centre and the Magellanic clouds.

\section{Summary and conclusions}
\label{sec:conclusions}

In this study, we have focused on the capability to detect unresolved binary systems in \Gaia DR3, using \ruwe as a key metric. We developed a forward model for predicting \ruwe values, which depend on the sky coordinates and a set of astrophysical properties of the observed stellar system (single or binary). This model allowed us to determine the upper limit of \ruwe that will describe good \Gaia measurements for single sources. Consequently, we established a \ruwe threshold that varies across different regions of the sky (Fig.~\ref{fig:ruwe_singleSource}), where \ruwe values exceeding this threshold will potentially indicate the presence of an unseen companion. Then, using the approach described in \citet{2023A&A...677A..37C} to estimate the selection function of subsamples of \Gaia data, we estimate the selection probability of a \Gaia source with a \ruwe above our sky-dependent threshold. We note that this selection probability will trace areas where binary systems can be more, or less, easily detected across most of the sky, except in the very crowded regions. In these regions, Fig~\ref{fig:binaries_SF} will trace contamination when selecting binary systems solely based on \ruwe, since this will be increased due to contributions other than binarity.

We simulate a broad range of binary systems to describe what types of systems, e.g., mass and luminosity ratios, period or distances, will lead to a \ruwe above the threshold and therefore make these systems detectable using this approach. The probability of detecting these systems decreases with increasing distance. Specifically, about $50\%$ of the simulated binary systems were detectable at a distance of approximately $100$ pc. However, for binary systems with orbital periods as long as the temporal baseline of the survey, this detectability range extends significantly, allowing for the detection of half of these systems at distances up to about $1$ kpc.  We also see that the imprints of binarity on \ruwe are weaker for systems with orbital periods around $1$ or $1/2$ years \citep[due to a confusion with a parallax shift,][]{2020MNRAS.495..321P} and mass ratios going to zero (single source) or one (no apparent wobbling of the photocentre around the CoM). Therefore, systems with these characteristics are less likely to be identified as binaries in \Gaia when relying exclusively on \ruwe criteria.

Finally, we use the GUMS simulation within $200$ pc and apply our forward model for \ruwe, to provide insight into the unresolved binary population in the GCNS. Our sky-varying \ruwe criteria for identifying potential binary systems outperforms the widely used threshold value of $1.4$, both in terms of the completeness and the purity of the resulting sample. In the GUMS sample, we find that the most numerous binary systems, when both components are MS stars, are not the easiest to detect in \Gaia. The stronger imprints in \ruwe, thus higher detectability, are generated by binary systems with a luminous primary component (high-MS, mid-MS or Giant star). On the other hand, binary systems with a dim primary component (WD or low-MS stars) will have a low detection probability. For instance, only about $1\%$ of WD-WD binary systems simulated in GUMS were detected.

This study provides a new definition of what imprints on \ruwe we can expect from binarity, and it is adapted to the systems' location on-sky. However, most of the existing binary systems in our Galaxy will still go unnoticed due to the CoM wobbling around the photocentre being dominated by the astrometric noise of \Gaia, therefore resulting in a \ruwe value below the threshold. The tools we have developed provide a valuable resource to detect binary system candidates in \Gaia data and characterise the properties of the population from which the selected systems are drawn. Following the example of previous GaiaUnlimited selection functions, we make the code and tools generated in this work publicly available through the GaiaUnlimited Python package.
\begin{acknowledgements}

We thank David W. Hogg for his contributions to the GaiaUnlimited project. We thank Douglas Boubert, Andrew Everall and Tereza Jerabkova for their help in preparing the GUMS binary catalogue.

This work is a result of the GaiaUnlimited project, which has
received funding from the European Union’s Horizon 2020
research and innovation program under grant agreement No
101004110. The GaiaUnlimited project was started at the 2019
Santa Barbara Gaia Sprint, hosted by the Kavli Institute for Theoretical Physics at the University of California, Santa Barbara.

ZP acknowledges that this project has received funding from the European Research Council (ERC) under the European Union’s Horizon 2020 research and innovation programme (Grant agreement No. 101002511 - VEGA P).

This work has made use of results from the European Space Agency (ESA)
space mission {\it Gaia}, the data from which were processed by the {\it Gaia
Data Processing and Analysis Consortium} (DPAC).  Funding for the DPAC
has been provided by national institutions, in particular, the
institutions participating in the {\it Gaia} Multilateral Agreement. The
{\it Gaia} mission website is \url{http: //www.cosmos.esa.int/gaia}. The
authors are current or past members of the ESA {\it Gaia} mission team and
of the {\it Gaia} DPAC.
This work made use of Astropy:\footnote{\url{http://www.astropy.org}} a community-developed core Python package and an ecosystem of tools and resources for astronomy \citep{2022ApJ...935..167A}.

\end{acknowledgements}

\bibliographystyle{aa}
\bibliography{bibliography}

\begin{thebibliography}{36}
\expandafter\ifx\csname natexlab\endcsname\relax\def\natexlab#1{#1}\fi

\bibitem[{{Andrew} {et~al.}(2022){Andrew}, {Penoyre}, {Belokurov}, {Evans}, \&
  {Oh}}]{2022MNRAS.516.3661A}
{Andrew}, S., {Penoyre}, Z., {Belokurov}, V., {Evans}, N.~W., \& {Oh}, S. 2022,
  \mnras, 516, 3661

\bibitem[{{Astropy Collaboration} {et~al.}(2022){Astropy Collaboration},
  {Price-Whelan}, {Lim}, {Earl}, {Starkman}, {Bradley}, {Shupe}, {Patil},
  {Corrales}, {Brasseur}, {N{\"o}the}, {Donath}, {Tollerud}, {Morris},
  {Ginsburg}, {Vaher}, {Weaver}, {Tocknell}, {Jamieson}, {van Kerkwijk},
  {Robitaille}, {Merry}, {Bachetti}, {G{\"u}nther}, {Aldcroft},
  {Alvarado-Montes}, {Archibald}, {B{\'o}di}, {Bapat}, {Barentsen},
  {Baz{\'a}n}, {Biswas}, {Boquien}, {Burke}, {Cara}, {Cara}, {Conroy},
  {Conseil}, {Craig}, {Cross}, {Cruz}, {D'Eugenio}, {Dencheva}, {Devillepoix},
  {Dietrich}, {Eigenbrot}, {Erben}, {Ferreira}, {Foreman-Mackey}, {Fox},
  {Freij}, {Garg}, {Geda}, {Glattly}, {Gondhalekar}, {Gordon}, {Grant},
  {Greenfield}, {Groener}, {Guest}, {Gurovich}, {Handberg}, {Hart},
  {Hatfield-Dodds}, {Homeier}, {Hosseinzadeh}, {Jenness}, {Jones}, {Joseph},
  {Kalmbach}, {Karamehmetoglu}, {Ka{\l}uszy{\'n}ski}, {Kelley}, {Kern},
  {Kerzendorf}, {Koch}, {Kulumani}, {Lee}, {Ly}, {Ma}, {MacBride}, {Maljaars},
  {Muna}, {Murphy}, {Norman}, {O'Steen}, {Oman}, {Pacifici}, {Pascual},
  {Pascual-Granado}, {Patil}, {Perren}, {Pickering}, {Rastogi}, {Roulston},
  {Ryan}, {Rykoff}, {Sabater}, {Sakurikar}, {Salgado}, {Sanghi}, {Saunders},
  {Savchenko}, {Schwardt}, {Seifert-Eckert}, {Shih}, {Jain}, {Shukla}, {Sick},
  {Simpson}, {Singanamalla}, {Singer}, {Singhal}, {Sinha}, {Sip{\H{o}}cz},
  {Spitler}, {Stansby}, {Streicher}, {{\v{S}}umak}, {Swinbank}, {Taranu},
  {Tewary}, {Tremblay}, {de Val-Borro}, {Van Kooten}, {Vasovi{\'c}}, {Verma},
  {de Miranda Cardoso}, {Williams}, {Wilson}, {Winkel}, {Wood-Vasey}, {Xue},
  {Yoachim}, {Zhang}, {Zonca}, \& {Astropy Project
  Contributors}}]{2022ApJ...935..167A}
{Astropy Collaboration}, {Price-Whelan}, A.~M., {Lim}, P.~L., {et~al.} 2022,
  \apj, 935, 167

\bibitem[{{Belokurov} {et~al.}(2020){Belokurov}, {Penoyre}, {Oh}, {Iorio},
  {Hodgkin}, {Evans}, {Everall}, {Koposov}, {Tout}, {Izzard}, {Clarke}, \&
  {Brown}}]{2020MNRAS.496.1922B}
{Belokurov}, V., {Penoyre}, Z., {Oh}, S., {et~al.} 2020, \mnras, 496, 1922

\bibitem[{{Bressan} {et~al.}(2012){Bressan}, {Marigo}, {Girardi}, {Salasnich},
  {Dal Cero}, {Rubele}, \& {Nanni}}]{2012MNRAS.427..127B}
{Bressan}, A., {Marigo}, P., {Girardi}, L., {et~al.} 2012, \mnras, 427, 127

\bibitem[{{Cantat-Gaudin} {et~al.}(2023){Cantat-Gaudin}, {Fouesneau}, {Rix},
  {Brown}, {Castro-Ginard}, {Kostrzewa-Rutkowska}, {Drimmel}, {Hogg}, {Casey},
  {Khanna}, {Oh}, {Price-Whelan}, {Belokurov}, {Saydjari}, \&
  {Green}}]{2023A&A...669A..55C}
{Cantat-Gaudin}, T., {Fouesneau}, M., {Rix}, H.-W., {et~al.} 2023, \aap, 669,
  A55

\bibitem[{{Cantat-Gaudin} {et~al.}(2024){Cantat-Gaudin}, {Fouesneau}, {Rix},
  {Brown}, {Drimmel}, {Castro-Ginard}, {Khanna}, {Belokurov}, \&
  {Casey}}]{2024arXiv240105023C}
{Cantat-Gaudin}, T., {Fouesneau}, M., {Rix}, H.-W., {et~al.} 2024, arXiv
  e-prints, arXiv:2401.05023

\bibitem[{{Castro-Ginard} {et~al.}(2023){Castro-Ginard}, {Brown},
  {Kostrzewa-Rutkowska}, {Cantat-Gaudin}, {Drimmel}, {Oh}, {Belokurov},
  {Casey}, {Fouesneau}, {Khanna}, {Price-Whelan}, \&
  {Rix}}]{2023A&A...677A..37C}
{Castro-Ginard}, A., {Brown}, A. G.~A., {Kostrzewa-Rutkowska}, Z., {et~al.}
  2023, \aap, 677, A37

\bibitem[{{Duch{\^e}ne} \& {Kraus}(2013)}]{2013ARA&A..51..269D}
{Duch{\^e}ne}, G. \& {Kraus}, A. 2013, \araa, 51, 269

\bibitem[{{Duch{\^e}ne} {et~al.}(2018){Duch{\^e}ne}, {Lacour}, {Moraux},
  {Goodwin}, \& {Bouvier}}]{2018MNRAS.478.1825D}
{Duch{\^e}ne}, G., {Lacour}, S., {Moraux}, E., {Goodwin}, S., \& {Bouvier}, J.
  2018, \mnras, 478, 1825

\bibitem[{{El-Badry}(2024)}]{2024arXiv240312146E}
{El-Badry}, K. 2024, arXiv e-prints, arXiv:2403.12146

\bibitem[{{El-Badry} {et~al.}(2021){El-Badry}, {Rix}, \&
  {Heintz}}]{2021MNRAS.506.2269E}
{El-Badry}, K., {Rix}, H.-W., \& {Heintz}, T.~M. 2021, \mnras, 506, 2269

\bibitem[{{Fern\'andez-Hern\'andez} {et~al.}(2022){Fern\'andez-Hern\'andez},
  {Joliet}, \& {Garralda}}]{GOST}
{Fern\'andez-Hern\'andez}, J., {Joliet}, E., \& {Garralda}, N. 2022, GOST
  Software User Manual,
  \url{https://gaia.esac.esa.int/gost/docs/gost_software_user_manual.pdf}

\bibitem[{{Fitton} {et~al.}(2022){Fitton}, {Tofflemire}, \&
  {Kraus}}]{2022RNAAS...6...18F}
{Fitton}, S., {Tofflemire}, B.~M., \& {Kraus}, A.~L. 2022, Research Notes of
  the American Astronomical Society, 6, 18

\bibitem[{{Gaia Collaboration} {et~al.}(2018){Gaia Collaboration}, {Brown},
  {Vallenari}, {Prusti}, {de Bruijne}, {Babusiaux}, {Bailer-Jones}, {Biermann},
  {Evans}, {Eyer}, \& et~al.}]{2018A&A...616A...1G}
{Gaia Collaboration}, {Brown}, A.~G.~A., {Vallenari}, A., {et~al.} 2018, \aap,
  616, A1

\bibitem[{{Gaia Collaboration} {et~al.}(2021{\natexlab{a}}){Gaia
  Collaboration}, {Brown}, {Vallenari}, {Prusti}, {de Bruijne}, {Babusiaux},
  {Biermann}, {Creevey}, {Evans}, {Eyer}, {Hutton}, {Jansen}, {Jordi},
  {Klioner}, {Lammers}, {Lindegren}, {Luri}, {Mignard}, {Panem}, {Pourbaix},
  {Randich}, {Sartoretti}, {Soubiran}, {Walton}, {Arenou}, {Bailer-Jones},
  {Bastian}, {Cropper}, {Drimmel}, {Katz}, {Lattanzi}, {van Leeuwen}, {Bakker},
  {Cacciari}, {Casta{\~n}eda}, {De Angeli}, {Ducourant}, {Fabricius},
  {Fouesneau}, {Fr{\'e}mat}, {Guerra}, {Guerrier}, {Guiraud}, {Jean-Antoine
  Piccolo}, {Masana}, {Messineo}, {Mowlavi}, {Nicolas}, {Nienartowicz},
  {Pailler}, {Panuzzo}, {Riclet}, {Roux}, {Seabroke}, {Sordo}, {Tanga},
  {Th{\'e}venin}, {Gracia-Abril}, {Portell}, {Teyssier}, {Altmann}, {Andrae},
  {Bellas-Velidis}, {Benson}, {Berthier}, {Blomme}, {Brugaletta}, {Burgess},
  {Busso}, {Carry}, {Cellino}, {Cheek}, {Clementini}, {Damerdji}, {Davidson},
  {Delchambre}, {Dell'Oro}, {Fern{\'a}ndez-Hern{\'a}ndez}, {Galluccio},
  {Garc{\'\i}a-Lario}, {Garcia-Reinaldos}, {Gonz{\'a}lez-N{\'u}{\~n}ez},
  {Gosset}, {Haigron}, {Halbwachs}, {Hambly}, {Harrison}, {Hatzidimitriou},
  {Heiter}, {Hern{\'a}ndez}, {Hestroffer}, {Hodgkin}, {Holl}, {Jan{\ss}en},
  {Jevardat de Fombelle}, {Jordan}, {Krone-Martins}, {Lanzafame},
  {L{\"o}ffler}, {Lorca}, {Manteiga}, {Marchal}, {Marrese}, {Moitinho}, {Mora},
  {Muinonen}, {Osborne}, {Pancino}, {Pauwels}, {Petit}, {Recio-Blanco},
  {Richards}, {Riello}, {Rimoldini}, {Robin}, {Roegiers}, {Rybizki}, {Sarro},
  {Siopis}, {Smith}, {Sozzetti}, {Ulla}, {Utrilla}, {van Leeuwen}, {van
  Reeven}, {Abbas}, {Abreu Aramburu}, {Accart}, {Aerts}, {Aguado}, {Ajaj},
  {Altavilla}, {{\'A}lvarez}, {{\'A}lvarez Cid-Fuentes}, {Alves}, {Anderson},
  {Anglada Varela}, {Antoja}, {Audard}, {Baines}, {Baker},
  {Balaguer-N{\'u}{\~n}ez}, {Balbinot}, {Balog}, {Barache}, {Barbato},
  {Barros}, {Barstow}, {Bartolom{\'e}}, {Bassilana}, {Bauchet},
  {Baudesson-Stella}, {Becciani}, {Bellazzini}, {Bernet}, {Bertone}, {Bianchi},
  {Blanco-Cuaresma}, {Boch}, {Bombrun}, {Bossini}, {Bouquillon}, {Bragaglia},
  {Bramante}, {Breedt}, {Bressan}, {Brouillet}, {Bucciarelli}, {Burlacu},
  {Busonero}, {Butkevich}, {Buzzi}, {Caffau}, {Cancelliere}, {C{\'a}novas},
  {Cantat-Gaudin}, {Carballo}, {Carlucci}, {Carnerero}, {Carrasco},
  {Casamiquela}, {Castellani}, {Castro-Ginard}, {Castro Sampol}, {Chaoul},
  {Charlot}, {Chemin}, {Chiavassa}, {Cioni}, {Comoretto}, {Cooper}, {Cornez},
  {Cowell}, {Crifo}, {Crosta}, {Crowley}, {Dafonte}, {Dapergolas}, {David},
  {David}, {de Laverny}, {De Luise}, {De March}, {De Ridder}, {de Souza}, {de
  Teodoro}, {de Torres}, {del Peloso}, {del Pozo}, {Delbo}, {Delgado},
  {Delgado}, {Delisle}, {Di Matteo}, {Diakite}, {Diener}, {Distefano},
  {Dolding}, {Eappachen}, {Edvardsson}, {Enke}, {Esquej}, {Fabre}, {Fabrizio},
  {Faigler}, {Fedorets}, {Fernique}, {Fienga}, {Figueras}, {Fouron},
  {Fragkoudi}, {Fraile}, {Franke}, {Gai}, {Garabato}, {Garcia-Gutierrez},
  {Garc{\'\i}a-Torres}, {Garofalo}, {Gavras}, {Gerlach}, {Geyer}, {Giacobbe},
  {Gilmore}, {Girona}, {Giuffrida}, {Gomel}, {Gomez}, {Gonzalez-Santamaria},
  {Gonz{\'a}lez-Vidal}, {Granvik}, {Guti{\'e}rrez-S{\'a}nchez}, {Guy},
  {Hauser}, {Haywood}, {Helmi}, {Hidalgo}, {Hilger}, {H{\l}adczuk}, {Hobbs},
  {Holland}, {Huckle}, {Jasniewicz}, {Jonker}, {Juaristi Campillo}, {Julbe},
  {Karbevska}, {Kervella}, {Khanna}, {Kochoska}, {Kontizas}, {Kordopatis},
  {Korn}, {Kostrzewa-Rutkowska}, {Kruszy{\'n}ska}, {Lambert}, {Lanza}, {Lasne},
  {Le Campion}, {Le Fustec}, {Lebreton}, {Lebzelter}, {Leccia}, {Leclerc},
  {Lecoeur-Taibi}, {Liao}, {Licata}, {Lindstr{\o}m}, {Lister}, {Livanou},
  {Lobel}, {Madrero Pardo}, {Managau}, {Mann}, {Marchant}, {Marconi}, {Marcos
  Santos}, {Marinoni}, {Marocco}, {Marshall}, {Martin Polo},
  {Mart{\'\i}n-Fleitas}, {Masip}, {Massari}, {Mastrobuono-Battisti}, {Mazeh},
  {McMillan}, {Messina}, {Michalik}, {Millar}, {Mints}, {Molina}, {Molinaro},
  {Moln{\'a}r}, {Montegriffo}, {Mor}, {Morbidelli}, {Morel}, {Morris},
  {Mulone}, {Munoz}, {Muraveva}, {Murphy}, {Musella}, {Noval}, {Ord{\'e}novic},
  {Orr{\`u}}, {Osinde}, {Pagani}, {Pagano}, {Palaversa}, {Palicio}, {Panahi},
  {Pawlak}, {Pe{\~n}alosa Esteller}, {Penttil{\"a}}, {Piersimoni}, {Pineau},
  {Plachy}, {Plum}, {Poggio}, {Poretti}, {Poujoulet}, {Pr{\v{s}}a}, {Pulone},
  {Racero}, {Ragaini}, {Rainer}, {Raiteri}, {Rambaux}, {Ramos}, {Ramos-Lerate},
  {Re Fiorentin}, {Regibo}, {Reyl{\'e}}, {Ripepi}, {Riva}, {Rixon}, {Robichon},
  {Robin}, {Roelens}, {Rohrbasser}, {Romero-G{\'o}mez}, {Rowell}, {Royer},
  {Rybicki}, {Sadowski}, {Sagrist{\`a} Sell{\'e}s}, {Sahlmann}, {Salgado},
  {Salguero}, {Samaras}, {Sanchez Gimenez}, {Sanna}, {Santove{\~n}a},
  {Sarasso}, {Schultheis}, {Sciacca}, {Segol}, {Segovia}, {S{\'e}gransan},
  {Semeux}, {Shahaf}, {Siddiqui}, {Siebert}, {Siltala}, {Slezak}, {Smart},
  {Solano}, {Solitro}, {Souami}, {Souchay}, {Spagna}, {Spoto}, {Steele},
  {Steidelm{\"u}ller}, {Stephenson}, {S{\"u}veges}, {Szabados}, {Szegedi-Elek},
  {Taris}, {Tauran}, {Taylor}, {Teixeira}, {Thuillot}, {Tonello}, {Torra},
  {Torra}, {Turon}, {Unger}, {Vaillant}, {van Dillen}, {Vanel}, {Vecchiato},
  {Viala}, {Vicente}, {Voutsinas}, {Weiler}, {Wevers}, {Wyrzykowski}, {Yoldas},
  {Yvard}, {Zhao}, {Zorec}, {Zucker}, {Zurbach}, \&
  {Zwitter}}]{2021A&A...649A...1G}
{Gaia Collaboration}, {Brown}, A.~G.~A., {Vallenari}, A., {et~al.}
  2021{\natexlab{a}}, \aap, 649, A1

\bibitem[{{Gaia Collaboration} {et~al.}(2016){Gaia Collaboration}, {Prusti},
  {de Bruijne}, {Brown}, {Vallenari}, {Babusiaux}, {Bailer-Jones}, {Bastian},
  {Biermann}, {Evans}, \& et~al.}]{2016A&A...595A...1G}
{Gaia Collaboration}, {Prusti}, T., {de Bruijne}, J.~H.~J., {et~al.} 2016,
  \aap, 595, A1

\bibitem[{{Gaia Collaboration} {et~al.}(2021{\natexlab{b}}){Gaia
  Collaboration}, {Smart}, {Sarro}, {Rybizki}, {Reyl{\'e}}, {Robin}, {Hambly},
  {Abbas}, {Barstow}, {de Bruijne}, {Bucciarelli}, {Carrasco}, {Cooper},
  {Hodgkin}, {Masana}, {Michalik}, {Sahlmann}, {Sozzetti}, {Brown},
  {Vallenari}, {Prusti}, {Babusiaux}, {Biermann}, {Creevey}, {Evans}, {Eyer},
  {Hutton}, {Jansen}, {Jordi}, {Klioner}, {Lammers}, {Lindegren}, {Luri},
  {Mignard}, {Panem}, {Pourbaix}, {Randich}, {Sartoretti}, {Soubiran},
  {Walton}, {Arenou}, {Bailer-Jones}, {Bastian}, {Cropper}, {Drimmel}, {Katz},
  {Lattanzi}, {van Leeuwen}, {Bakker}, {Casta{\~n}eda}, {De Angeli},
  {Ducourant}, {Fabricius}, {Fouesneau}, {Fr{\'e}mat}, {Guerra}, {Guerrier},
  {Guiraud}, {Jean-Antoine Piccolo}, {Messineo}, {Mowlavi}, {Nicolas},
  {Nienartowicz}, {Pailler}, {Panuzzo}, {Riclet}, {Roux}, {Seabroke}, {Sordo},
  {Tanga}, {Th{\'e}venin}, {Gracia-Abril}, {Portell}, {Teyssier}, {Altmann},
  {Andrae}, {Bellas-Velidis}, {Benson}, {Berthier}, {Blomme}, {Brugaletta},
  {Burgess}, {Busso}, {Carry}, {Cellino}, {Cheek}, {Clementini}, {Damerdji},
  {Davidson}, {Delchambre}, {Dell'Oro}, {Fern{\'a}ndez-Hern{\'a}ndez},
  {Galluccio}, {Garc{\'\i}a-Lario}, {Garcia-Reinaldos},
  {Gonz{\'a}lez-N{\'u}{\~n}ez}, {Gosset}, {Haigron}, {Halbwachs}, {Harrison},
  {Hatzidimitriou}, {Heiter}, {Hern{\'a}ndez}, {Hestroffer}, {Holl},
  {Jan{\ss}en}, {Jevardat de Fombelle}, {Jordan}, {Krone-Martins}, {Lanzafame},
  {L{\"o}ffler}, {Lorca}, {Manteiga}, {Marchal}, {Marrese}, {Moitinho}, {Mora},
  {Muinonen}, {Osborne}, {Pancino}, {Pauwels}, {Recio-Blanco}, {Richards},
  {Riello}, {Rimoldini}, {Roegiers}, {Siopis}, {Smith}, {Ulla}, {Utrilla}, {van
  Leeuwen}, {van Reeven}, {Abreu Aramburu}, {Accart}, {Aerts}, {Aguado},
  {Ajaj}, {Altavilla}, {{\'A}lvarez}, {{\'A}lvarez Cid-Fuentes}, {Alves},
  {Anderson}, {Anglada Varela}, {Antoja}, {Audard}, {Baines}, {Baker},
  {Balaguer-N{\'u}{\~n}ez}, {Balbinot}, {Balog}, {Barache}, {Barbato},
  {Barros}, {Bartolom{\'e}}, {Bassilana}, {Bauchet}, {Baudesson-Stella},
  {Becciani}, {Bellazzini}, {Bernet}, {Bertone}, {Bianchi}, {Blanco-Cuaresma},
  {Boch}, {Bombrun}, {Bossini}, {Bouquillon}, {Bragaglia}, {Bramante},
  {Breedt}, {Bressan}, {Brouillet}, {Burlacu}, {Busonero}, {Butkevich},
  {Buzzi}, {Caffau}, {Cancelliere}, {C{\'a}novas}, {Cantat-Gaudin}, {Carballo},
  {Carlucci}, {Carnerero}, {Casamiquela}, {Castellani}, {Castro-Ginard},
  {Castro Sampol}, {Chaoul}, {Charlot}, {Chemin}, {Chiavassa}, {Cioni},
  {Comoretto}, {Cornez}, {Cowell}, {Crifo}, {Crosta}, {Crowley}, {Dafonte},
  {Dapergolas}, {David}, {David}, {de Laverny}, {De Luise}, {De March}, {De
  Ridder}, {de Souza}, {de Teodoro}, {de Torres}, {del Peloso}, {del Pozo},
  {Delgado}, {Delgado}, {Delisle}, {Di Matteo}, {Diakite}, {Diener},
  {Distefano}, {Dolding}, {Eappachen}, {Edvardsson}, {Enke}, {Esquej}, {Fabre},
  {Fabrizio}, {Faigler}, {Fedorets}, {Fernique}, {Fienga}, {Figueras},
  {Fouron}, {Fragkoudi}, {Fraile}, {Franke}, {Gai}, {Garabato},
  {Garcia-Gutierrez}, {Garc{\'\i}a-Torres}, {Garofalo}, {Gavras}, {Gerlach},
  {Geyer}, {Giacobbe}, {Gilmore}, {Girona}, {Giuffrida}, {Gomel}, {Gomez},
  {Gonzalez-Santamaria}, {Gonz{\'a}lez-Vidal}, {Granvik},
  {Guti{\'e}rrez-S{\'a}nchez}, {Guy}, {Hauser}, {Haywood}, {Helmi}, {Hidalgo},
  {Hilger}, {H{\l}adczuk}, {Hobbs}, {Holland}, {Huckle}, {Jasniewicz},
  {Jonker}, {Juaristi Campillo}, {Julbe}, {Karbevska}, {Kervella}, {Khanna},
  {Kochoska}, {Kontizas}, {Kordopatis}, {Korn}, {Kostrzewa-Rutkowska},
  {Kruszy{\'n}ska}, {Lambert}, {Lanza}, {Lasne}, {Le Campion}, {Le Fustec},
  {Lebreton}, {Lebzelter}, {Leccia}, {Leclerc}, {Lecoeur-Taibi}, {Liao},
  {Licata}, {Lindstr{\o}m}, {Lister}, {Livanou}, {Lobel}, {Madrero Pardo},
  {Managau}, {Mann}, {Marchant}, {Marconi}, {Marcos Santos}, {Marinoni},
  {Marocco}, {Marshall}, {Martin Polo}, {Mart{\'\i}n-Fleitas}, {Masip},
  {Massari}, {Mastrobuono-Battisti}, {Mazeh}, {McMillan}, {Messina}, {Millar},
  {Mints}, {Molina}, {Molinaro}, {Moln{\'a}r}, {Montegriffo}, {Mor},
  {Morbidelli}, {Morel}, {Morris}, {Mulone}, {Munoz}, {Muraveva}, {Murphy},
  {Musella}, {Noval}, {Ord{\'e}novic}, {Orr{\`u}}, {Osinde}, {Pagani},
  {Pagano}, {Palaversa}, {Palicio}, {Panahi}, {Pawlak}, {Pe{\~n}alosa
  Esteller}, {Penttil{\"a}}, {Piersimoni}, {Pineau}, {Plachy}, {Plum},
  {Poggio}, {Poretti}, {Poujoulet}, {Pr{\v{s}}a}, {Pulone}, {Racero},
  {Ragaini}, {Rainer}, {Raiteri}, {Rambaux}, {Ramos}, {Ramos-Lerate}, {Re
  Fiorentin}, {Regibo}, {Ripepi}, {Riva}, {Rixon}, {Robichon}, {Robin},
  {Roelens}, {Rohrbasser}, {Romero-G{\'o}mez}, {Rowell}, {Royer}, {Rybicki},
  {Sadowski}, {Sagrist{\`a} Sell{\'e}s}, {Salgado}, {Salguero}, {Samaras},
  {Sanchez Gimenez}, {Sanna}, {Santove{\~n}a}, {Sarasso}, {Schultheis},
  {Sciacca}, {Segol}, {Segovia}, {S{\'e}gransan}, {Semeux}, {Shahaf},
  {Siddiqui}, {Siebert}, {Siltala}, {Slezak}, {Solano}, {Solitro}, {Souami},
  {Souchay}, {Spagna}, {Spoto}, {Steele}, {Steidelm{\"u}ller}, {Stephenson},
  {S{\"u}veges}, {Szabados}, {Szegedi-Elek}, {Taris}, {Tauran}, {Taylor},
  {Teixeira}, {Thuillot}, {Tonello}, {Torra}, {Torra}, {Turon}, {Unger},
  {Vaillant}, {van Dillen}, {Vanel}, {Vecchiato}, {Viala}, {Vicente},
  {Voutsinas}, {Weiler}, {Wevers}, {Wyrzykowski}, {Yoldas}, {Yvard}, {Zhao},
  {Zorec}, {Zucker}, {Zurbach}, \& {Zwitter}}]{2021A&A...649A...6G}
{Gaia Collaboration}, {Smart}, R.~L., {Sarro}, L.~M., {et~al.}
  2021{\natexlab{b}}, \aap, 649, A6

\bibitem[{{Gaia Collaboration} {et~al.}(2023){Gaia Collaboration}, {Vallenari},
  {Brown}, {Prusti}, {de Bruijne}, {Arenou}, {Babusiaux}, {Biermann},
  {Creevey}, {Ducourant}, {Evans}, {Eyer}, {Guerra}, {Hutton}, {Jordi},
  {Klioner}, {Lammers}, {Lindegren}, {Luri}, {Mignard}, {Panem}, {Pourbaix},
  {Randich}, {Sartoretti}, {Soubiran}, {Tanga}, {Walton}, {Bailer-Jones},
  {Bastian}, {Drimmel}, {Jansen}, {Katz}, {Lattanzi}, {van Leeuwen}, {Bakker},
  {Cacciari}, {Casta{\~n}eda}, {De Angeli}, {Fabricius}, {Fouesneau},
  {Fr{\'e}mat}, {Galluccio}, {Guerrier}, {Heiter}, {Masana}, {Messineo},
  {Mowlavi}, {Nicolas}, {Nienartowicz}, {Pailler}, {Panuzzo}, {Riclet}, {Roux},
  {Seabroke}, {Sordo}, {Th{\'e}venin}, {Gracia-Abril}, {Portell}, {Teyssier},
  {Altmann}, {Andrae}, {Audard}, {Bellas-Velidis}, {Benson}, {Berthier},
  {Blomme}, {Burgess}, {Busonero}, {Busso}, {C{\'a}novas}, {Carry}, {Cellino},
  {Cheek}, {Clementini}, {Damerdji}, {Davidson}, {de Teodoro}, {Nu{\~n}ez
  Campos}, {Delchambre}, {Dell'Oro}, {Esquej}, {Fern{\'a}ndez-Hern{\'a}ndez},
  {Fraile}, {Garabato}, {Garc{\'\i}a-Lario}, {Gosset}, {Haigron}, {Halbwachs},
  {Hambly}, {Harrison}, {Hern{\'a}ndez}, {Hestroffer}, {Hodgkin}, {Holl},
  {Jan{\ss}en}, {Jevardat de Fombelle}, {Jordan}, {Krone-Martins}, {Lanzafame},
  {L{\"o}ffler}, {Marchal}, {Marrese}, {Moitinho}, {Muinonen}, {Osborne},
  {Pancino}, {Pauwels}, {Recio-Blanco}, {Reyl{\'e}}, {Riello}, {Rimoldini},
  {Roegiers}, {Rybizki}, {Sarro}, {Siopis}, {Smith}, {Sozzetti}, {Utrilla},
  {van Leeuwen}, {Abbas}, {{\'A}brah{\'a}m}, {Abreu Aramburu}, {Aerts},
  {Aguado}, {Ajaj}, {Aldea-Montero}, {Altavilla}, {{\'A}lvarez}, {Alves},
  {Anders}, {Anderson}, {Anglada Varela}, {Antoja}, {Baines}, {Baker},
  {Balaguer-N{\'u}{\~n}ez}, {Balbinot}, {Balog}, {Barache}, {Barbato},
  {Barros}, {Barstow}, {Bartolom{\'e}}, {Bassilana}, {Bauchet}, {Becciani},
  {Bellazzini}, {Berihuete}, {Bernet}, {Bertone}, {Bianchi}, {Binnenfeld},
  {Blanco-Cuaresma}, {Blazere}, {Boch}, {Bombrun}, {Bossini}, {Bouquillon},
  {Bragaglia}, {Bramante}, {Breedt}, {Bressan}, {Brouillet}, {Brugaletta},
  {Bucciarelli}, {Burlacu}, {Butkevich}, {Buzzi}, {Caffau}, {Cancelliere},
  {Cantat-Gaudin}, {Carballo}, {Carlucci}, {Carnerero}, {Carrasco},
  {Casamiquela}, {Castellani}, {Castro-Ginard}, {Chaoul}, {Charlot}, {Chemin},
  {Chiaramida}, {Chiavassa}, {Chornay}, {Comoretto}, {Contursi}, {Cooper},
  {Cornez}, {Cowell}, {Crifo}, {Cropper}, {Crosta}, {Crowley}, {Dafonte},
  {Dapergolas}, {David}, {David}, {de Laverny}, {De Luise}, {De March}, {De
  Ridder}, {de Souza}, {de Torres}, {del Peloso}, {del Pozo}, {Delbo},
  {Delgado}, {Delisle}, {Demouchy}, {Dharmawardena}, {Di Matteo}, {Diakite},
  {Diener}, {Distefano}, {Dolding}, {Edvardsson}, {Enke}, {Fabre}, {Fabrizio},
  {Faigler}, {Fedorets}, {Fernique}, {Fienga}, {Figueras}, {Fournier},
  {Fouron}, {Fragkoudi}, {Gai}, {Garcia-Gutierrez}, {Garcia-Reinaldos},
  {Garc{\'\i}a-Torres}, {Garofalo}, {Gavel}, {Gavras}, {Gerlach}, {Geyer},
  {Giacobbe}, {Gilmore}, {Girona}, {Giuffrida}, {Gomel}, {Gomez},
  {Gonz{\'a}lez-N{\'u}{\~n}ez}, {Gonz{\'a}lez-Santamar{\'\i}a},
  {Gonz{\'a}lez-Vidal}, {Granvik}, {Guillout}, {Guiraud},
  {Guti{\'e}rrez-S{\'a}nchez}, {Guy}, {Hatzidimitriou}, {Hauser}, {Haywood},
  {Helmer}, {Helmi}, {Sarmiento}, {Hidalgo}, {Hilger}, {H{\l}adczuk}, {Hobbs},
  {Holland}, {Huckle}, {Jardine}, {Jasniewicz}, {Jean-Antoine Piccolo},
  {Jim{\'e}nez-Arranz}, {Jorissen}, {Juaristi Campillo}, {Julbe}, {Karbevska},
  {Kervella}, {Khanna}, {Kontizas}, {Kordopatis}, {Korn}, {K{\'o}sp{\'a}l},
  {Kostrzewa-Rutkowska}, {Kruszy{\'n}ska}, {Kun}, {Laizeau}, {Lambert},
  {Lanza}, {Lasne}, {Le Campion}, {Lebreton}, {Lebzelter}, {Leccia}, {Leclerc},
  {Lecoeur-Taibi}, {Liao}, {Licata}, {Lindstr{\o}m}, {Lister}, {Livanou},
  {Lobel}, {Lorca}, {Loup}, {Madrero Pardo}, {Magdaleno Romeo}, {Managau},
  {Mann}, {Manteiga}, {Marchant}, {Marconi}, {Marcos}, {Marcos Santos},
  {Mar{\'\i}n Pina}, {Marinoni}, {Marocco}, {Marshall}, {Martin Polo},
  {Mart{\'\i}n-Fleitas}, {Marton}, {Mary}, {Masip}, {Massari},
  {Mastrobuono-Battisti}, {Mazeh}, {McMillan}, {Messina}, {Michalik}, {Millar},
  {Mints}, {Molina}, {Molinaro}, {Moln{\'a}r}, {Monari}, {Mongui{\'o}},
  {Montegriffo}, {Montero}, {Mor}, {Mora}, {Morbidelli}, {Morel}, {Morris},
  {Muraveva}, {Murphy}, {Musella}, {Nagy}, {Noval}, {Oca{\~n}a}, {Ogden},
  {Ordenovic}, {Osinde}, {Pagani}, {Pagano}, {Palaversa}, {Palicio},
  {Pallas-Quintela}, {Panahi}, {Payne-Wardenaar}, {Pe{\~n}alosa Esteller},
  {Penttil{\"a}}, {Pichon}, {Piersimoni}, {Pineau}, {Plachy}, {Plum}, {Poggio},
  {Pr{\v{s}}a}, {Pulone}, {Racero}, {Ragaini}, {Rainer}, {Raiteri}, {Rambaux},
  {Ramos}, {Ramos-Lerate}, {Re Fiorentin}, {Regibo}, {Richards}, {Rios Diaz},
  {Ripepi}, {Riva}, {Rix}, {Rixon}, {Robichon}, {Robin}, {Robin}, {Roelens},
  {Rogues}, {Rohrbasser}, {Romero-G{\'o}mez}, {Rowell}, {Royer}, {Ruz Mieres},
  {Rybicki}, {Sadowski}, {S{\'a}ez N{\'u}{\~n}ez}, {Sagrist{\`a} Sell{\'e}s},
  {Sahlmann}, {Salguero}, {Samaras}, {Sanchez Gimenez}, {Sanna},
  {Santove{\~n}a}, {Sarasso}, {Schultheis}, {Sciacca}, {Segol}, {Segovia},
  {S{\'e}gransan}, {Semeux}, {Shahaf}, {Siddiqui}, {Siebert}, {Siltala},
  {Silvelo}, {Slezak}, {Slezak}, {Smart}, {Snaith}, {Solano}, {Solitro},
  {Souami}, {Souchay}, {Spagna}, {Spina}, {Spoto}, {Steele},
  {Steidelm{\"u}ller}, {Stephenson}, {S{\"u}veges}, {Surdej}, {Szabados},
  {Szegedi-Elek}, {Taris}, {Taylor}, {Teixeira}, {Tolomei}, {Tonello}, {Torra},
  {Torra}, {Torralba Elipe}, {Trabucchi}, {Tsounis}, {Turon}, {Ulla}, {Unger},
  {Vaillant}, {van Dillen}, {van Reeven}, {Vanel}, {Vecchiato}, {Viala},
  {Vicente}, {Voutsinas}, {Weiler}, {Wevers}, {Wyrzykowski}, {Yoldas}, {Yvard},
  {Zhao}, {Zorec}, {Zucker}, \& {Zwitter}}]{2023A&A...674A...1G}
{Gaia Collaboration}, {Vallenari}, A., {Brown}, A.~G.~A., {et~al.} 2023, \aap,
  674, A1

\bibitem[{{Halbwachs} {et~al.}(2023){Halbwachs}, {Pourbaix}, {Arenou},
  {Galluccio}, {Guillout}, {Bauchet}, {Marchal}, {Sadowski}, \&
  {Teyssier}}]{2023A&A...674A...9H}
{Halbwachs}, J.-L., {Pourbaix}, D., {Arenou}, F., {et~al.} 2023, \aap, 674, A9

\bibitem[{{Kervella} {et~al.}(2022){Kervella}, {Arenou}, \&
  {Th{\'e}venin}}]{2022A&A...657A...7K}
{Kervella}, P., {Arenou}, F., \& {Th{\'e}venin}, F. 2022, \aap, 657, A7

\bibitem[{Lindegren(2018)}]{LL:LL-124}
Lindegren, L. 2018, gAIA-C3-TN-LU-LL-124

\bibitem[{Lindegren(2022)}]{LL:LL-136}
Lindegren, L. 2022, gAIA-C3-TN-LU-LL-136

\bibitem[{{Lindegren} {et~al.}(2021{\natexlab{a}}){Lindegren}, {Bastian},
  {Biermann}, {Bombrun}, {de Torres}, {Gerlach}, {Geyer}, {Hern{\'a}ndez},
  {Hilger}, {Hobbs}, {Klioner}, {Lammers}, {McMillan}, {Ramos-Lerate},
  {Steidelm{\"u}ller}, {Stephenson}, \& {van Leeuwen}}]{2021A&A...649A...4L}
{Lindegren}, L., {Bastian}, U., {Biermann}, M., {et~al.} 2021{\natexlab{a}},
  \aap, 649, A4

\bibitem[{{Lindegren} {et~al.}(2021{\natexlab{b}}){Lindegren}, {Klioner},
  {Hern{\'a}ndez}, {Bombrun}, {Ramos-Lerate}, {Steidelm{\"u}ller}, {Bastian},
  {Biermann}, {de Torres}, {Gerlach}, {Geyer}, {Hilger}, {Hobbs}, {Lammers},
  {McMillan}, {Stephenson}, {Casta{\~n}eda}, {Davidson}, {Fabricius},
  {Gracia-Abril}, {Portell}, {Rowell}, {Teyssier}, {Torra}, {Bartolom{\'e}},
  {Clotet}, {Garralda}, {Gonz{\'a}lez-Vidal}, {Torra}, {Abbas}, {Altmann},
  {Anglada Varela}, {Balaguer-N{\'u}{\~n}ez}, {Balog}, {Barache}, {Becciani},
  {Bernet}, {Bertone}, {Bianchi}, {Bouquillon}, {Brown}, {Bucciarelli},
  {Busonero}, {Butkevich}, {Buzzi}, {Cancelliere}, {Carlucci}, {Charlot},
  {Cioni}, {Crosta}, {Crowley}, {del Peloso}, {del Pozo}, {Drimmel}, {Esquej},
  {Fienga}, {Fraile}, {Gai}, {Garcia-Reinaldos}, {Guerra}, {Hambly}, {Hauser},
  {Jan{\ss}en}, {Jordan}, {Kostrzewa-Rutkowska}, {Lattanzi}, {Liao}, {Licata},
  {Lister}, {L{\"o}ffler}, {Marchant}, {Masip}, {Mignard}, {Mints}, {Molina},
  {Mora}, {Morbidelli}, {Murphy}, {Pagani}, {Panuzzo}, {Pe{\~n}alosa Esteller},
  {Poggio}, {Re Fiorentin}, {Riva}, {Sagrist{\`a} Sell{\'e}s}, {Sanchez
  Gimenez}, {Sarasso}, {Sciacca}, {Siddiqui}, {Smart}, {Souami}, {Spagna},
  {Steele}, {Taris}, {Utrilla}, {van Reeven}, \&
  {Vecchiato}}]{2021A&A...649A...2L}
{Lindegren}, L., {Klioner}, S.~A., {Hern{\'a}ndez}, J., {et~al.}
  2021{\natexlab{b}}, \aap, 649, A2

\bibitem[{{Luri, X.} {et~al.}(2014){Luri, X.}, {Palmer, M.}, {Arenou, F.},
  {Masana, E.}, {de Bruijne, J.}, {Antiche, E.}, {Babusiaux, C.}, {Borrachero,
  R.}, {Sartoretti, P.}, {Julbe, F.}, {Isasi, Y.}, {Martinez, O.}, {Robin, A.
  C.}, {Reylé, C.}, {Jordi, C.}, \& {Carrasco, J. M.}}]{gog}
{Luri, X.}, {Palmer, M.}, {Arenou, F.}, {et~al.} 2014, \aap, 566, A119

\bibitem[{{Majewski} {et~al.}(2017){Majewski}, {Schiavon}, {Frinchaboy},
  {Allende Prieto}, {Barkhouser}, {Bizyaev}, {Blank}, {Brunner}, {Burton},
  {Carrera}, {Chojnowski}, {Cunha}, {Epstein}, {Fitzgerald}, {Garc{\'\i}a
  P{\'e}rez}, {Hearty}, {Henderson}, {Holtzman}, {Johnson}, {Lam}, {Lawler},
  {Maseman}, {M{\'e}sz{\'a}ros}, {Nelson}, {Nguyen}, {Nidever}, {Pinsonneault},
  {Shetrone}, {Smee}, {Smith}, {Stolberg}, {Skrutskie}, {Walker}, {Wilson},
  {Zasowski}, {Anders}, {Basu}, {Beland}, {Blanton}, {Bovy}, {Brownstein},
  {Carlberg}, {Chaplin}, {Chiappini}, {Eisenstein}, {Elsworth}, {Feuillet},
  {Fleming}, {Galbraith-Frew}, {Garc{\'\i}a}, {Garc{\'\i}a-Hern{\'a}ndez},
  {Gillespie}, {Girardi}, {Gunn}, {Hasselquist}, {Hayden}, {Hekker}, {Ivans},
  {Kinemuchi}, {Klaene}, {Mahadevan}, {Mathur}, {Mosser}, {Muna}, {Munn},
  {Nichol}, {O'Connell}, {Parejko}, {Robin}, {Rocha-Pinto}, {Schultheis},
  {Serenelli}, {Shane}, {Silva Aguirre}, {Sobeck}, {Thompson}, {Troup},
  {Weinberg}, \& {Zamora}}]{2017AJ....154...94M}
{Majewski}, S.~R., {Schiavon}, R.~P., {Frinchaboy}, P.~M., {et~al.} 2017, \aj,
  154, 94

\bibitem[{{Penoyre} {et~al.}(2022{\natexlab{a}}){Penoyre}, {Belokurov}, \&
  {Evans}}]{2022MNRAS.513.2437P}
{Penoyre}, Z., {Belokurov}, V., \& {Evans}, N.~W. 2022{\natexlab{a}}, \mnras,
  513, 2437

\bibitem[{{Penoyre} {et~al.}(2022{\natexlab{b}}){Penoyre}, {Belokurov}, \&
  {Evans}}]{2022MNRAS.513.5270P}
{Penoyre}, Z., {Belokurov}, V., \& {Evans}, N.~W. 2022{\natexlab{b}}, \mnras,
  513, 5270

\bibitem[{{Penoyre} {et~al.}(2020){Penoyre}, {Belokurov}, {Wyn Evans},
  {Everall}, \& {Koposov}}]{2020MNRAS.495..321P}
{Penoyre}, Z., {Belokurov}, V., {Wyn Evans}, N., {Everall}, A., \& {Koposov},
  S.~E. 2020, \mnras, 495, 321

\bibitem[{{Price-Whelan} {et~al.}(2020){Price-Whelan}, {Hogg}, {Rix}, {Beaton},
  {Lewis}, {Nidever}, {Almeida}, {Badenes}, {Barba}, {Beers}, {Carlberg}, {De
  Lee}, {Fern{\'a}ndez-Trincado}, {Frinchaboy}, {Garc{\'\i}a-Hern{\'a}ndez},
  {Green}, {Hasselquist}, {Longa-Pe{\~n}a}, {Majewski}, {Nitschelm}, {Sobeck},
  {Stassun}, {Stringfellow}, \& {Troup}}]{2020ApJ...895....2P}
{Price-Whelan}, A.~M., {Hogg}, D.~W., {Rix}, H.-W., {et~al.} 2020, \apj, 895, 2

\bibitem[{{Raghavan} {et~al.}(2010){Raghavan}, {McAlister}, {Henry}, {Latham},
  {Marcy}, {Mason}, {Gies}, {White}, \& {ten Brummelaar}}]{2010ApJS..190....1R}
{Raghavan}, D., {McAlister}, H.~A., {Henry}, T.~J., {et~al.} 2010, \apjs, 190,
  1

\bibitem[{{Robin} {et~al.}(2012){Robin}, {Luri}, {Reyl{\'e}}, {Isasi}, {Grux},
  {Blanco-Cuaresma}, {Arenou}, {Babusiaux}, {Belcheva}, {Drimmel}, {Jordi},
  {Krone-Martins}, {Masana}, {Mauduit}, {Mignard}, {Mowlavi},
  {Rocca-Volmerange}, {Sartoretti}, {Slezak}, \&
  {Sozzetti}}]{2012A&A...543A.100R}
{Robin}, A.~C., {Luri}, X., {Reyl{\'e}}, C., {et~al.} 2012, \aap, 543, A100

\bibitem[{{Skrutskie} {et~al.}(2006){Skrutskie}, {Cutri}, {Stiening},
  {Weinberg}, {Schneider}, {Carpenter}, {Beichman}, {Capps}, {Chester},
  {Elias}, {Huchra}, {Liebert}, {Lonsdale}, {Monet}, {Price}, {Seitzer},
  {Jarrett}, {Kirkpatrick}, {Gizis}, {Howard}, {Evans}, {Fowler}, {Fullmer},
  {Hurt}, {Light}, {Kopan}, {Marsh}, {McCallon}, {Tam}, {Van Dyk}, \&
  {Wheelock}}]{2mass}
{Skrutskie}, M.~F., {Cutri}, R.~M., {Stiening}, R., {et~al.} 2006, \aj, 131,
  1163

\bibitem[{{Stassun} \& {Torres}(2021)}]{2021ApJ...907L..33S}
{Stassun}, K.~G. \& {Torres}, G. 2021, \apjl, 907, L33

\bibitem[{{Wallace}(2024)}]{2024MNRAS.527.8718W}
{Wallace}, A.~L. 2024, \mnras, 527, 8718

\bibitem[{{Wright} {et~al.}(2010){Wright}, {Eisenhardt}, {Mainzer}, {Ressler},
  {Cutri}, {Jarrett}, {Kirkpatrick}, {Padgett}, {McMillan}, {Skrutskie},
  {Stanford}, {Cohen}, {Walker}, {Mather}, {Leisawitz}, {Gautier}, {McLean},
  {Benford}, {Lonsdale}, {Blain}, {Mendez}, {Irace}, {Duval}, {Liu}, {Royer},
  {Heinrichsen}, {Howard}, {Shannon}, {Kendall}, {Walsh}, {Larsen}, {Cardon},
  {Schick}, {Schwalm}, {Abid}, {Fabinsky}, {Naes}, \&
  {Tsai}}]{2010AJ....140.1868W}
{Wright}, E.~L., {Eisenhardt}, P. R.~M., {Mainzer}, A.~K., {et~al.} 2010, \aj,
  140, 1868

\end{thebibliography}

\onecolumn
\begin{appendix}
\section{Using the GaiaUnlimited python package}
\label{sec:gaiaunlimited}

We provide the code we used to estimate the selection function of binary systems as a new tool in the GaiaUnlimited python package\footnote{\url{https://github.com/gaia-unlimited/gaiaunlimited}}. The necessary data, i.e., all-sky scanning law details at a resolution of HEALPix level 5, to generate the plots in the paper are also included within the package.

Listing~\ref{lst:figures} shows how to use the new \texttt{BinarySystemsSelectionFunction} class. Similarly to the previous GaiaUnlimited selection functions, it takes an \texttt{astropy} coordinate object and returns the queried quantity, e.g., sky-varying \ruwe threshold (to produce Fig~\ref{fig:ruwe_singleSource}) or selection probabilities of sources with high \ruwe \citep[computed as in][produces Fig.~\ref{fig:binaries_SF}]{2023A&A...677A..37C} or the \Gaia scanning law details.

\begin{minipage}{0.95\columnwidth}
\begin{lstlisting}[language=Python, caption={Example of code to initialise and query the selection function of binary systems.},label={lst:figures}]
from gaiaunlimited.selectionfunctions import binaries

SF = binaries.BinarySystemsSelectionFunction()

from astropy import units as u
from astropy.coordinates import SkyCoord

coord = SkyCoord(ra = ra*u.degree, dec = dec*u.degree, frame = 'icrs')

# Generates the selection probabilities based on RUWE. 
# Function to generate Fig. 4.
probability,variance = SF.query(coord, return_variance = True)

# Returns the RUWE threshold above which a source is considered a potential binary system. 
# Function to generate Fig. 3.
ruwe = SF.query_RUWE(coord,crowding = True)
\end{lstlisting}
\end{minipage}

We also provide our forward model to compute a \ruwe value from the binary system properties, and the \Gaia scanning law in the \texttt{SimulateGaiaSource} class. Listing~\ref{lst:simulation} shows how to initialise the simulation of a system at a given sky location, providing the period, eccentricity and initial phase of the system, as well as the \Gaia scanning law details (observation times, scanning angles and parallax factors, provided by GOST). The AL measurements and errors are computed as a function of the other system properties. Finally, a \ruwe value is computed from these simulated observations. Full documentation of the package is available through the GaiaUnlimited documentation website\footnote{\url{https://gaiaunlimited.readthedocs.io/en/latest/}}.

\begin{minipage}{0.95\columnwidth}
\begin{lstlisting}[language=Python, caption={Example of code to simulate a \Gaia observation as a function of the source parameters (single or binary), and estimate its \ruwe.},label={lst:simulation}]
from gaiaunlimited.utils import SimulateGaiaSource

source = SimulateGaiaSource(ra, dec, period, eccentricity, initial_phase)

al_pos,al_err = source.observe(Gmag,
                        parallax,
                        semimajor_axis,
                        mass_ratio,
                        luminosity_ratio,
                        phi,
                        theta,
                        omega,)

ruwe = source.unit_weight_error(al_pos, al_err)
\end{lstlisting}
\end{minipage}

\end{appendix}
\end{document}